\documentclass[conference,compsoc]{IEEEtran}

\ifCLASSOPTIONcompsoc
\usepackage[nocompress]{cite}
\else
\usepackage{cite}
\fi
\usepackage{url}

\ifCLASSINFOpdf
\else
\fi

\ifCLASSOPTIONcompsoc
\usepackage[caption=false,font=footnotesize,labelfont=sf,textfont=sf]{subfig}
\else
\usepackage[caption=false,font=footnotesize]{subfig}
\fi

\usepackage{tikz}
\usepackage{bbding}
\usepackage{graphicx}
\usepackage{xcolor}
\usepackage{booktabs}
\usepackage{tabularx}
\usepackage{multirow}
\usepackage{tcolorbox}
\usepackage{algpseudocode}
\usepackage{listings}
\usepackage{enumitem}
\usepackage{xspace}
\usepackage[textwidth=0.65in,textsize=scriptsize,colorinlistoftodos]{todonotes}
\let\pkgtodo\todo

\ifdefined\cryptofigpreambleloaded\else
\def\cryptofigpreambleloaded{}

\makeatletter
\@ifpackageloaded{tikz}{}{\usepackage{tikz}}
\makeatother

\usetikzlibrary{
  calc,
  positioning,
  fit,
  arrows.meta,
  backgrounds,
  shadows.blur,
  shapes.callouts,
  shapes.geometric,
  shapes.symbols,
  decorations.pathmorphing,
  decorations.markings,
  decorations.pathreplacing
}

\ifdefined\cryptofigcolorsloaded\else
\def\cryptofigcolorsloaded{}

\definecolor{sitebg}{HTML}{EAEAEA}
\definecolor{cardbg}{HTML}{D9D9D9}
\definecolor{stepboxbg}{HTML}{EBEBEB}

\definecolor{icongreen}{HTML}{8CC08C}
\definecolor{iconred}{HTML}{FF8282}
\definecolor{malboxbg}{HTML}{FFCFCF}
\definecolor{malboxborder}{HTML}{FFB9B9}
\definecolor{attackred}{HTML}{FF0000}
\definecolor{leakblue}{HTML}{08A2F6}
\definecolor{stargold}{HTML}{F5A623}

\fi

\providecommand{\figdebug}[2][]{}

\tikzset{
  fig shadow/.style={
    blur shadow={
      shadow blur steps=6,
      shadow xshift=0.4mm,
      shadow yshift=-0.4mm,
      shadow opacity=20,
    },
  },
}

\ifdefined\cryptofiglayoutloaded\else
\def\cryptofiglayoutloaded{}

\tikzset{
  fig panel/.style={
    draw=black!20,
    fill=black!5,
    rounded corners=8pt,
    line width=0.8pt,
  },
  fig danger box/.style={
    draw=malboxborder,
    fill=malboxbg,
    thick,
    rounded corners=2pt,
    font=\footnotesize\itshape,
    inner sep=3pt,
    anchor=west,
  },
  fig leak icon/.style={
    inner sep=2pt,
    fill=leakblue!35,
    rounded corners=2pt,
  },
}

\fi
 
\fi
 \definecolor{mGreen}{rgb}{0,0.6,0}
\definecolor{mGray}{rgb}{0.5,0.5,0.5}
\definecolor{mPurple}{rgb}{0.58,0,0.82}
\definecolor{backgroundColour}{rgb}{1,1,1}
\definecolor{lightgreen}{rgb}{0.8,1.0,0.8}
\definecolor{reviewgreen}{rgb}{0.55,0.75,0.55}
\definecolor{darkgreen}{rgb}{0.0,0.5,0.0}
\definecolor{lightblue}{rgb}{0.68,0.85,0.9}
\definecolor{pastelcyan}{rgb}{0.62,0.78,0.86}
\definecolor{lightpurple}{rgb}{0.7,0.55,0.88}

\definecolor{diffstart}{named}{blue}
\definecolor{diffrem}{named}{red}
\definecolor{black}{named}{black}

\lstdefinestyle{CStyleSmall}{
  backgroundcolor=\color{backgroundColour},
  commentstyle=\color{mGreen},
  keywordstyle=\color{mGray},
  numberstyle=\tiny\color{mGray},
  stringstyle=\color{black},
  basicstyle=\scriptsize\sffamily,
  breakatwhitespace=false,
  breaklines=true,
  captionpos=b,
  keepspaces=true,
  numbers=left,
  numbersep=5pt,
  showspaces=false,
  showstringspaces=false,
  showtabs=false,
  tabsize=2,
  breakindent=0pt,
  moredelim=**[is][\color{diffrem}]{\\-}{\\-},
  moredelim=**[is][\color{darkgreen}]{\\+}{\\+},
  moredelim=**[is][\color{mGray}]{\\gray}{\\gray},
  moredelim=**[is][\color{mPurple}]{\\purple}{\\purple},
  moredelim=**[is][\bfseries]{\\b}{\\b},
  moredelim=**[is][\mdseries]{\\ub}{\\ub},
  moredelim=**[is][\color{black}]{\\uc}{\\uc},
}

\lstdefinestyle{diff}{
  basicstyle=\footnotesize\ttfamily,
  backgroundcolor=\color{backgroundColour},
  breakatwhitespace=false,
  breaklines=true,
  captionpos=b,
  keepspaces=true,
  numbers=left,
  numbersep=5pt,
  showspaces=false,
  showstringspaces=false,
  showtabs=false,
  tabsize=2,
  morecomment=[f][\color{diffstart}]{@@},
  morecomment=[f][\color{darkgreen}]{+},
  morecomment=[f][\color{diffrem}]{-},
}

\newlist{compactenum}{enumerate}{1}
\setlist[compactenum,1]{label=\arabic*.,nosep}

\newlist{compactlist}{itemize}{3}
\setlist[compactlist]{label=\textbullet,nosep}

\newif\ifpresentMode
\presentModefalse

\ifpresentMode
\renewcommand{\todo}[1]{}
\else
\renewcommand{\todo}[1]{\noindent\textcolor{red}{TODO: #1} \\}
\fi

\ifpresentMode
\newcommand{\mnote}[2][]{}
\newcommand{\mnoteyellow}[1]{}
\newcommand{\mnotered}[1]{}
\else
\newcommand{\mnote}[2][yellow]{\pkgtodo[color=#1!40,size=\scriptsize,linecolor=#1!60]{#2}}
\newcommand{\mnoteyellow}[1]{\mnote[yellow]{#1}}
\newcommand{\mnotered}[1]{\mnote[red]{#1}}
\fi

\ifpresentMode
\newcommand{\cv}[1]{}
\newcommand{\sk}[1]{}
\newcommand{\ac}[1]{}
\newcommand{\raluca}[1]{}

\else
\newcommand{\cv}[1]{\noindent\textcolor{blue}{Corban: #1}}
\newcommand{\sk}[1]{\noindent\textcolor{mPurple}{Sohee: #1}}
\newcommand{\ac}[1]{\noindent\textcolor{lightpurple}{Austin: #1}}
\newcommand{\raluca}[1]{\noindent\textcolor{lightblue}{Raluca: #1}}

\fi

\ifpresentMode
\newcommand{\old}[1]{}

\else
\newcommand{\old}[1]{{\color{red}#1}}

\fi

\newcommand{\sys}{Chai\xspace}
\newcommand{\Bugs}{\mathsf{Bugs}}

\tcbset{
  takeaway/.style={
    colback=gray!10, colframe=gray!60, coltext=black, boxrule=0.5pt, arc=5pt, left=5pt, right=5pt, top=2pt, bottom=2pt, boxsep=2pt, grow to left by=-10pt,
    grow to right by=-10pt,
  },
}

 \newcommand{\legdot}[1]{\protect\textcolor{#1}{\protect\raisebox{-0.45ex}{\scalebox{1.265}{\large$\bullet$}}}}
\newcommand{\legendapp}{\legdot{teal!40}}
\newcommand{\legendlib}{\legdot{violet!35}}
\newcommand{\legendspec}{\legdot{orange!50}}

\begin{document}
\title{Chai: Agentic Discovery of Cryptographic Misuse Vulnerabilities}

\author{\IEEEauthorblockN{Corban Villa, Sohee Kim, Austin Chu, Alon Shakevsky, Raluca Ada Popa}
\IEEEauthorblockA{UC Berkeley}}

\maketitle

\begin{abstract}
  AI-assisted vulnerability discovery has proven effective for bug classes like memory
safety, where instrumentation confirms memory violations and efficiently filters false
positives. Many dangerous vulnerability classes, such as cryptographic misuse, however,
lack any comparable instrumentation. In this work, we present Chai, an AI-based system
that discovers and validates cryptographic misuse vulnerabilities through naturally
occurring signals. To achieve this, Chai rethinks the classical technique of
differential testing by leveraging AI to 1) improve precision for detecting real security
issues in libraries, and 2) repurpose commonly overlooked discrepancies as leads for
tangible vulnerabilities in downstream applications. In doing so, Chai inverts the
prevailing paradigm of AI vulnerability discovery: instead of auditing one codebase for
many flaws, it catalogs flaws at the library level and propagates them across a
cryptographic dependency graph, delivering compounding efficiency gains. We evaluate
Chai across X.509, JWT, and SAML libraries. Chai discovered a previously unknown
critical vulnerability in an SSL library that powers billions of devices, along with
security bugs in one library behind a major web browser and another in major Linux
distributions. In total, these techniques surfaced over 100 vulnerabilities.
 \end{abstract}

\IEEEpeerreviewmaketitle

\section{Introduction}

Building secure software requires not just sound design but continuous effort to find flaws before they are exploited. Developers search for vulnerabilities using techniques from static analysis and fuzzing to end-to-end penetration testing. Historically, defenders have benefited from the scarce expertise needed to discover and exploit vulnerabilities, limiting adversaries' ability to operate at scale~\cite{bilgeWeKnewIt2012,finifterEmpiricalStudyVulnerability2013,herleyNobodySellsGold2010}. AI-assisted vulnerability discovery is challenging this assumption: introducing models into the pipeline, whether augmenting classical techniques or driving end-to-end approaches such as Anthropic's Mythos~\cite{anthropicMythosPreview2026}, has improved results across vulnerability classes and surfaced flaws of increasing complexity~\cite{wolfsslClaudeMythosHarden2026,wangCyberGymEvaluatingAI2025a,wangExploitGymCanAI2026,leeExploitBenchCapabilityLadder2026,schneier2026instantsoftware}. CVE disclosures in the most recent 90-day window grew ${\sim}$50\% year-over-year~\cite{nistNvdCveApi}, while reported time-to-exploit fell from ${\sim}$21.5 days in 2025 to 24 hours~\cite{zeroDayClock2026}.

As the cost of discovery and exploitation falls, the vulnerability classes most likely to be weaponized demand attention. Yet many systems concentrate on verifiable domains, where fuzzing infrastructure, crash reproduction, and sanitizers provide oracles that confirm findings and filter hallucinated or deceptive results~\cite{serebryanyAddressSanitizerFastAddress2012,googleOssFuzzAiPowered2023,googleOssFuzzLevelingUp2024,googleBigSleep2024,anthropicMythosPreview2026,zhangAixccSok2026,chinOssCrs2026,shengFuzzingBrainV22026}. These systems have produced findings in high volume, yet likely exploitable
  memory-safety vulnerabilities occurred roughly 3.4$\times$ less frequently
  than a year prior~\cite{berkeleyVulnInitiative}. This shows that volume alone does not
  translate into real-world impact. By contrast, {\em{cryptographic and authentication failures}} rank highly in both the OWASP Top Ten~\cite{owaspTopTen2021} and the MITRE Top 25~\cite{mitreTop252024}, and can propagate to thousands of downstream applications~\cite{durumericMatterHeartbleed2014,aviramDROWNBreakingTLS2016,bockReturnBleichenbachersOracle2018,blessingCryptographyWildEmpirical2024}. Making agent-based discovery practical here thus requires an address-sanitizer analogue, but one that validates higher-level security properties.

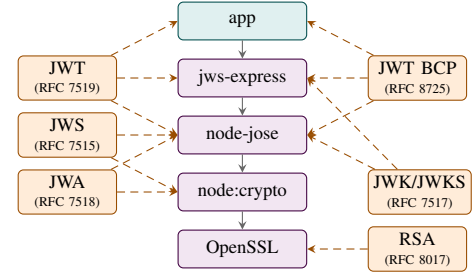
\begin{figure}[t]
  \centering
  \begin{tikzpicture}[
      node distance=2.4mm and 4mm,
      every node/.style={
        font=\scriptsize,
        rounded corners=2pt,
        minimum height=5mm,
        inner sep=2pt,
        align=center,
      },
      lib/.style={
        fill=violet!10,
        draw=violet!50!black,
        minimum width=17mm,
      },
      app/.style={
        fill=teal!12,
        draw=teal!60!black,
        minimum width=17mm,
      },
      spec/.style={
        fill=orange!15,
        draw=orange!60!black,
        minimum width=13mm,
      },
      arr/.style={
        ->,
        >=stealth,
        black!60,
      },
      specarr/.style={
        ->,
        >=stealth,
        orange!60!black,
        densely dashed,
      },
    ]

\node[app] (app) {app};
    \node[lib, below=of app] (middleware) {jws-express};
    \node[lib, below=of middleware] (nodejose) {node-jose};
    \node[lib, below=of nodejose] (nodecrypto) {node:crypto};
    \node[lib, below=of nodecrypto] (openssl) {OpenSSL};

\draw[arr] (app) -- (middleware);
    \draw[arr] (middleware) -- (nodejose);
    \draw[arr] (nodejose) -- (nodecrypto);
    \draw[arr] (nodecrypto) -- (openssl);

\node[spec, left=8mm of middleware] (jwt) {JWT\\[-1pt]\tiny(RFC 7519)};
    \draw[specarr] (jwt) -- (app.west);
    \draw[specarr] (jwt) -- (middleware.west);
    \draw[specarr] (jwt) -- (nodejose.west);

\node[spec, left=8mm of nodejose] (jws) {JWS\\[-1pt]\tiny(RFC 7515)};
    \draw[specarr] (jws) -- (nodejose.west);
    \draw[specarr] (jws) -- (nodecrypto.west);

\node[spec, left=8mm of nodecrypto] (jwa) {JWA\\[-1pt]\tiny(RFC 7518)};
    \draw[specarr] (jwa) -- (nodejose.west);
    \draw[specarr] (jwa) -- (nodecrypto.west);

\node[spec, right=8mm of middleware] (bcp) {JWT BCP\\[-1pt]\tiny(RFC 8725)};
    \draw[specarr] (bcp) -- (app.east);
    \draw[specarr] (bcp) -- (middleware.east);
    \draw[specarr] (bcp) -- (nodejose.east);

\node[spec, right=8mm of nodecrypto] (jwk) {JWK/JWKS\\[-1pt]\tiny(RFC 7517)};
    \draw[specarr] (jwk) -- (middleware.east);
    \draw[specarr] (jwk) -- (nodejose.east);

\node[spec, right=8mm of openssl] (rsa) {RSA\\[-1pt]\tiny(RFC 8017)};
    \draw[specarr] (rsa) -- (openssl);

  \end{tikzpicture}
  \caption{Cryptographic dependency stack for a JWT-verifying web
    application~(\legendapp), its
    underlying libraries~(\legendlib), and corresponding
  specifications~(\legendspec).}
  \label{fig:crypto-stack}
\end{figure}
 
Unfortunately, identifying cryptographic misuse is non-trivial. Consider a widely used
cryptographic authentication protocol such as JSON Web Tokens
(JWT). What appears to be a single protocol is in fact a composition of several
specifications for signing, encryption, algorithms, and key handling, realized
independently across dozens of libraries and languages (Figure~\ref{fig:crypto-stack}).
These implementations resolve the specification's under-specified corners differently, and
the consequences can be severe. For example, in 2018, the node-jose library accepted a
verification key embedded in the token itself by default, letting an attacker supply their
own key and forge valid tokens (CVE-2018-0114)~\cite{nvdCve20180114}. In isolation the behavior looked correct:
the signature verified, and the standard does permit embedding keys. Instead, there was a
mismatch in assumptions: the library assumed the application would decide which keys to
trust, while applications misinterpreted successful verification as trusted verification.
Whether the behavior is even wrong is not always clear-cut: these judgments are
subjective and nuanced, and can differ from one developer to the next for the same
protocol. There is no
address-sanitizer equivalent to confirm a violation or to signal what is correct. The
agent can argue that a vulnerability exists, but with no programmatic evidence to ground
the claim, security experts must filter through these issues one by one.

A long-standing approach to uncovering these types of issues in cryptographic libraries is
differential
testing, a classical technique introduced by McKeeman in 1998~\cite{mckeeman1998differential}. By running the same input
through multiple implementations and comparing their behavior, it surfaces disagreements
without needing to know the correct answer in advance. On its own, however, differential
testing is often noisy and still labor-intensive: it produces many non-actionable
discrepancies, and
steering it toward meaningful inputs has required protocol-specific grammars or mutators
that take considerable effort to build for each new protocol~\cite{brubakerUsingFrankencertsAutomated2014,petsiosNEZHAEfficientDomainIndependent2017,parachaHallucinatingCertificatesDifferential2026a,yangTokenTimeBomb2026}. Even then, for a complex
protocol the input space is large enough that exploring it thoroughly is non-trivial.

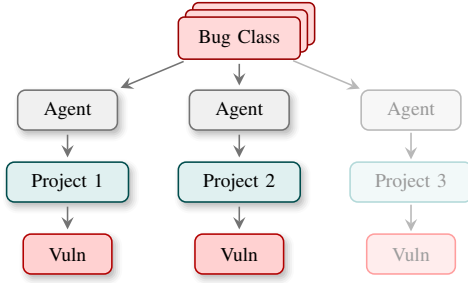
\begin{figure}[t]
  \centering
  \resizebox{0.72\columnwidth}{!}{\begin{tikzpicture}[
        font=\tiny,
        >=stealth,
        dropback/.style={blur shadow={shadow blur steps=4, shadow xshift=0.3mm, shadow
        yshift=-0.4mm, shadow opacity=30}},
        every node/.style={rounded corners=2pt, minimum height=3.6mm, inner sep=2pt, align=center},
        agent/.style={fill=black!6, draw=black!55, minimum width=9mm, dropback},
        gagent/.style={fill=black!3, draw=black!28, minimum width=9mm, text=black!40},
        proj/.style={fill=teal!12, draw=teal!60!black, minimum width=11mm, dropback},
        gproj/.style={fill=teal!5, draw=teal!30, minimum width=11mm, text=black!40},
        vmark/.style={fill=red!18, draw=red!70!black, text=black, minimum width=8mm,
        minimum height=3.6mm, inner sep=1pt, dropback},
        gvmark/.style={fill=red!7, draw=red!38, text=black!38, minimum width=8mm, minimum
        height=3.6mm, inner sep=1pt},
        bcard/.style={fill=red!15, draw=red!60!black, text=black, minimum width=11mm,
        minimum height=3.8mm, rounded corners=1.5pt, inner sep=1pt},
        arr/.style={->, black!55, shorten >=1pt, shorten <=1pt},
        farr/.style={->, black!45, thin, shorten >=0.5pt, shorten <=0.5pt},
        garr/.style={->, black!28, thin, shorten >=0.5pt, shorten <=0.5pt},
      ]
      \node[bcard] at (0.13,1.43) {};
      \node[bcard] at (0.065,1.365) {};
      \node[bcard] (bdeck) at (0,1.3) {Bug Class};
      \node[agent]  (ag1) at (-1.55,0.65) {Agent};
      \node[agent]  (ag2) at ( 0.0,0.65) {Agent};
      \node[gagent] (ag3) at ( 1.55,0.65) {Agent};
      \draw[arr] (bdeck) -- (ag1);
      \draw[arr] (bdeck) -- (ag2);
      \draw[garr] (bdeck) -- (ag3);
      \node[proj]  (p1) at (-1.55,0) {Project 1};
      \node[proj]  (p2) at ( 0.0,0) {Project 2};
      \node[gproj] (p3) at ( 1.55,0) {Project 3};
      \draw[arr] (ag1) -- (p1);
      \draw[arr] (ag2) -- (p2);
      \draw[garr] (ag3) -- (p3);
      \node[vmark]  (v1) at (-1.55,-0.65) {Vuln};
      \node[vmark]  (v2) at ( 0.0,-0.65) {Vuln};
      \node[gvmark] (v3) at ( 1.55,-0.65) {Vuln};
      \draw[arr] (p1) -- (v1);
      \draw[arr] (p2) -- (v2);
      \draw[garr] (p3) -- (v3);
    \end{tikzpicture}}
  \caption{\textbf{The prevailing paradigm.} Agents are often pointed at a project and
  tasked to search it for specific bug classes over many runs.}
  \label{fig:amplify}
\end{figure}

In this work, we present \sys, an AI-based system for discovering cryptographic misuse
vulnerabilities. Our first insight is to pair differential testing with AI so that each
covers the other's weakness, yielding agentic discovery with a signal we can check
rather than trusting the AI alone: differential testing supplies a verifiable and
deterministic signal that a discrepancy exists, while AI provides protocol knowledge and
adaptive search to surface the meaningful issues without hand-built grammars. This
pairing feeds our second insight, \emph{discrepancy tracing}, where we use ambiguities to
invert the prevailing paradigm of agentic vulnerability discovery
~\cite{anthropicMythosPreview2026,openaiCodexSecurityPreview2026,zhangAixccSok2026}
(Figure~\ref{fig:amplify}). Rather than start from a project
and search it for any number of potential bugs, we start from a concrete discrepancy,
surfaced by differential testing, and trace it to the projects it affects. Each discrepancy
resolves either to a flaw in the library itself or to an ambiguity that prior approaches
dismissed as non-reportable but that, we observe, often leads to tangible vulnerabilities
in downstream applications. In summary, \sys contributes a systematic approach to
discovering cryptographic misuse vulnerabilities through differential testing and AI.

To better understand the design motivation, consider our example for the JWT stack in
Figure~\ref{fig:crypto-stack}. A naive agent audits one application at a time, scanning
its codebase for bugs. These ambiguity-induced vulnerabilities, however, are often
invisible from a project's own codebase, since the code for the unintuitive behavior
lives in an external dependency: the application can look correct in isolation, and the
hazard surfaces only in the contract between it and the library beneath it. An agent
could, in principle, widen its scope to the dependencies, but the dependency tree quickly
explodes, and even then it re-derives a very similar analysis for every project built on
the same libraries, since the only part unique to each application is the slice of the
JWT, JWS, JWA, and JWK specifications its libraries leave unimplemented.

\sys proceeds in the opposite direction, scrutinizing the libraries first and working its
way up. It crafts a single input, such as the attacker-provided key token from before
(CVE-2018-0114), and runs it through every JWT implementation at once; wherever one
accepts what the others reject, the disagreement is recorded as a discrepancy. An agent
first triages these discrepancies, acting as a low-pass filter, and sorts each into a
library vulnerability, an ambiguity, or a non-actionable difference. Library
vulnerabilities and ambiguities are confirmed by manual review, and each ambiguity then
drives a reverse search. \sys traces it through the dependency graph to the applications
that depend on the library and directs an agent to audit each for that exact
behavior, here whether it trusts a key supplied in the token.
Where the straw man re-runs a full specification audit per project, \sys tests once and
follows each discrepancy to the applications it affects, compounding efficiency:
\begin{compactenum}
\item \textbf{Amplified testing.} A single input exercises every implementation of a
  protocol at once, rather than re-auditing each application against the same
  specification. For example, an agent execution produces a candidate that is tested
  against 20+ JWT libraries at once.
\item \textbf{Reverse search.} Each ambiguity is traced through the dependency graph to
  only the applications that could inherit it, rather than inspecting every project in
  turn. For example, a single JWT ambiguity may reach hundreds or thousands of dependent
  projects~\cite{zimmermannSmallWorldHigh2019}.
\item \textbf{Targeted auditing.} An agent audits each affected application for one
  specific behavior, rather than for any number of vulnerability classes or occurrences.
  For example, it checks only whether an application trusts a key supplied in the token.
\end{compactenum}

We implement \sys and evaluate it across X.509, JWT, and SAML, spanning 47 libraries across 8 languages. It surfaces several confirmed vulnerabilities and security-bug findings in major cryptographic libraries, along with over 100 under-investigation vulnerability findings in downstream applications. \sys's differential search compares favorably with prior differential-testing and fuzzing tools, surfacing ${\sim}2\times$ as many unique differences as the strongest baseline, many of which the baselines do not surface. It also finds a broader range of disagreements, including the hardest-to-reach cases where only a few libraries accept an input the rest reject, the strongest signal of a library-level flaw. Among the flaws it surfaced are two severe vulnerabilities in wolfSSL, a widely used cryptography library, both confirmed and patched by maintainers within hours of submission, along with security bugs in one library shipped in major Linux distributions and another library behind a major web browser. The wolfSSL codebase had recently been audited by Anthropic's Mythos~\cite{anthropicMythosPreview2026,wolfsslClaudeMythosHarden2026}, yet neither issue was surfaced, suggesting that \sys reaches vulnerabilities that many prevailing approaches overlook.
 \begin{figure*}[t]
  \tikzset{
  efficiency fig/.style={
    font=\scriptsize,
    >=stealth,
    dropback/.style={
      blur shadow={
        shadow blur steps=4,
        shadow xshift=0.3mm,
        shadow yshift=-0.4mm,
        shadow opacity=30,
      },
    },
    every node/.style={
      rounded corners=2pt,
      minimum height=4.6mm,
      minimum width=16mm,
      inner sep=2pt,
      align=center,
    },
    agent/.style={fill=black!6, draw=black!55, dropback},
    input/.style={fill=black!4, draw=black!45, dropback},
    proj/.style={fill=teal!12, draw=teal!60!black, dropback},
    smallproj/.style={fill=teal!12, draw=teal!60!black, minimum width=5.5mm, dropback},
    lib/.style={fill=violet!10, draw=violet!50!black, dropback},
    data/.style={fill=blue!8, draw=blue!55!black, dropback},
    ambig/.style={fill=orange!15, draw=orange!60!black, dropback},
    bug/.style={
      fill=red!16,
      draw=red!65!black,
      text=red!75!black,
      inner sep=1pt,
      dropback,
    },
    vmark/.style={
      fill=red!18,
      draw=red!70!black,
      text=red!75!black,
      inner sep=1pt,
      dropback,
    },
    dots/.style={
      draw=none,
      fill=none,
      font=\normalsize,
      text=black!55,
      inner sep=1pt,
      minimum width=4mm,
    },
    space/.style={draw=black!40, fill=black!7, rounded corners=3pt, inner sep=2.5pt},
    grouplab/.style={font=\scriptsize\itshape, text=black!60, anchor=south west, inner sep=0pt},
    arr/.style={->, black!55, shorten >=1.5pt, shorten <=1.5pt},
    farr/.style={->, black!45, thin, shorten >=1pt, shorten <=1pt},
    title/.style={font=\footnotesize\bfseries},
  },
}
   \footnotesize
  \noindent\makebox[\textwidth]{\begin{tabular}[c]{@{}c@{}}
      \refstepcounter{subfigure}\label{fig:efficiency-amplified-testing}\textbf{(\alph{subfigure}) Amplified testing}\\[1.5ex]
      \resizebox{!}{1.28in}{\begin{tikzpicture}[efficiency fig]
  \path[use as bounding box] (-2.78,-3.45) rectangle (2.05,0.25);
  \node[agent] (a1)   at (0,0) {Agent};
  \node[input] (i1)   at (0,-1.0) {Test Input};
  \node[lib]   (l1a)  at (-1.7,-2.0) {OpenSSL};
  \node[lib]   (l1b)  at ( 0.0,-2.0) {wolfSSL};
  \node[dots]  (l1c)  at ( 1.7,-2.0) {$\cdots$};
  \node[bug]   (bug1) at (-1.0,-3.05) {Bug};
  \node[ambig] (amb1) at ( 0.95,-3.05) {Ambiguity};
  \draw[arr] (a1) -- (i1);
  \draw[arr] (i1) -- (l1a);
  \draw[arr] (i1) -- (l1b);
  \draw[arr] (i1) -- (l1c);
  \draw[farr] (l1a) -- (bug1);
  \draw[farr] (l1b) -- (bug1);
  \draw[farr] (l1b) -- (amb1);
  \draw[farr] (l1c) -- (amb1);

  \begin{scope}[on background layer]
    \node[space, fit=(l1a)(l1b)(l1c)] (libgrp) {};
    \node[space, fit=(bug1)(amb1)] (outgrp) {};
  \end{scope}
  \node[grouplab] at ([xshift=-4pt,yshift=-1.5pt]libgrp.north west) {Libraries};
  \node[grouplab] at ([xshift=-4pt,yshift=-1.5pt]outgrp.north west) {Outputs};
\end{tikzpicture}
 }
    \end{tabular}\hspace{1.5em}\begin{tabular}[c]{@{}c@{}}
      \refstepcounter{subfigure}\label{fig:efficiency-reverse-search}\textbf{(\alph{subfigure}) Reverse search}\\[1.5ex]
      \resizebox{!}{1.28in}{\begin{tikzpicture}[efficiency fig]
  \path[use as bounding box] (-2.15,-3.45) rectangle (2.15,0.25);
  \node[ambig] (am2)  at (0,0) {Ambiguity};
  \node[lib]   (lib2) at (0,-1.0) {GnuTLS};
  \node[data]  (db2)  at (0,-2.0) {Database};
  \node[smallproj, minimum width=7mm] (p2a) at (-1.05,-3.0) {curl};
  \node[smallproj, minimum width=7mm] (p2b) at ( 0.0,-3.0) {wget};
  \node[dots]      (p2d) at ( 1.05,-3.0) {$\cdots$};
  \draw[arr] (am2) -- (lib2);
  \draw[arr] (lib2) -- (db2);
  \draw[farr] (db2) -- (p2a);
  \draw[farr] (db2) -- (p2b);
  \draw[farr] (db2) -- (p2d);

  \begin{scope}[on background layer]
    \node[space, fit=(p2a)(p2b)(p2d)] (appgrp) {};
  \end{scope}
  \node[grouplab] at ([xshift=-8pt,yshift=-1.5pt]appgrp.north west) {Apps};
\end{tikzpicture}
 }
    \end{tabular}\hspace{1.5em}\begin{tabular}[c]{@{}c@{}}
      \refstepcounter{subfigure}\label{fig:efficiency-targeted-audit}\textbf{(\alph{subfigure}) Targeted audit}\\[1.5ex]
      \resizebox{!}{1.28in}{\begin{tikzpicture}[efficiency fig]
  \path[use as bounding box] (-2.15,-3.45) rectangle (2.15,0.25);
  \node[ambig] (am3)  at (-1.05,0) {Ambiguity};
  \node[proj]  (app3) at ( 1.05,0) {App};
  \node[agent] (a3)   at (0,-1.05) {Agent};
  \node[vmark] (v3)   at (0,-2.1) {Vuln};
  \node[data]  (r3)   at (0,-3.05) {Report};
  \draw[arr] (am3)  -- (a3);
  \draw[arr] (app3) -- (a3);
  \draw[arr] (a3)   -- (v3);
  \draw[arr] (v3)   -- (r3);
\end{tikzpicture}
 }
    \end{tabular}\hfill
    \begin{minipage}[c]{0.20\textwidth}
      \caption{\textbf{\sys's compounding efficiency.} (a) A single input is
        amplified across all harnessed libraries, yielding bugs and ambiguities.
        (b) Each ambiguity is traced down the dependency graph to the apps that
        inherit it. (c) Each app receives a narrow, close-ended audit rather than
      an open-ended search.}
      \label{fig:efficiency}
    \end{minipage}}
\end{figure*}
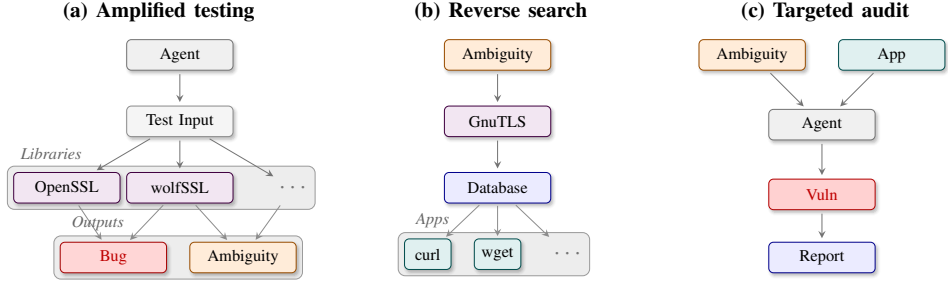
 \section{System Overview}

This section defines the threat model that \sys operates under, walks its
two-stage approach at a high level, then details the efficiency improvements
each stage delivers.

\subsection{Threat Model and Vulnerability Scope}
\label{sec:threat-model}

\sys identifies vulnerabilities in concrete implementations of cryptographic
protocols such as X.509, JWT, and SAML, where a standardized message can be
submitted to any conforming library and processing yields a security-relevant
decision: whether a signature is valid, whether a token is accepted, whether
access is granted. \sys does not directly detect flaws in the underlying
mathematical constructions or in the specifications themselves. An issue
surfaces only when at least two independent implementations disagree on the
same input.

We classify each behavioral discrepancy into one of four classes, assuming an
attacker who crafts and submits protocol messages, such as a certificate,
token, or assertion, to a verifying endpoint, but who holds no signing keys and
no privileged position in the deployment:

\begin{compactenum}
\item \textbf{Vulnerability.} This attacker can subvert a security decision as
  deployed.
\item \textbf{Security bug.} The library's behavior violates the specification,
  yet exploitation would require capabilities beyond this attacker, such as a
  particular configuration or a partially trusted party.
\item \textbf{Ambiguity.} The specification grants latitude or a maintainer
  makes an intentional design decision, leaving each library defensible in
  isolation. An ambiguity becomes a downstream vulnerability when an
  application relies on a security property its library does not enforce but
  another implementation would have. Some resolve instead to library
  vulnerabilities, fixed to improve misuse resistance: CVE-2018-0114 made
  token-embedded keys opt-in~\cite{nvdCve20180114}.
\item \textbf{Non-actionable difference.} The disagreement carries no
  security-relevant consequence at any layer.
\end{compactenum}

\sys's coverage is empirical rather than exhaustive. Its agents generate inputs
probabilistically, and the discrepancies they uncover represent a subset of
those present, concentrated in the behavior classes the agents are directed
toward. We evaluate this coverage empirically, measuring which classes of
issues \sys discovers and at what rate (\textsection\ref{sec:evaluation}).

\subsection{\sys's Approach}

\sys decomposes discovery into two independent stages, joined by a single
artifact: the discrepancy. This decomposition delivers compounding efficiency
gains over naive AI-driven approaches for these classes of cryptographic
vulnerabilities. In the first stage, an agent submits test inputs to every
harnessed implementation of a protocol at once and records each disagreement.
This stage uncovers vulnerabilities and security bugs in the libraries
themselves, while also surfacing the ambiguities that seed the second.

The second stage, discrepancy tracing, determines what each ambiguity, a
discrepancy identified as a missing guarantee, means downstream: an agent audits the
applications that
depend on the disagreeing library for reliance on that guarantee, yielding
application-level findings with proof-of-concept scripts.
Sections~\ref{sec:amplification} and~\ref{sec:tracing} detail each
stage's efficiency improvements; Section~\ref{sec:design} presents the
mechanisms.

\subsection{\sys's Amplified Testing}
\label{sec:amplification}

Many current approaches to AI-driven vulnerability discovery operate within a
single project at a time, scanning a codebase file-by-file~\cite{anthropicMythosPreview2026} or auditing its history
commit-by-commit~\cite{openaiCodexSecurityPreview2026}. At the scale of real software, this is unrealistic and
expensive: OpenSSL alone spans more than 3{,}400 source files and nearly 40{,}000
commits~\cite{opensslGithub}, beyond what any file-by-file or commit-by-commit sweep can
cover at reasonable cost. Historical commit searches carry a further penalty: the deeper into the history a
candidate flaw lies, the less likely it is to survive into the current release, so
effort spent on old commits increasingly yields findings that cannot be reproduced
on the versions anyone deploys.

Consider the space of defects that can arise in implementing a standard such as
X.509, written $\Bugs_{\mathsf{X.509}}$: every way an implementation might mishandle
the structures, checks, and semantics the specification prescribes. Each library
realizes its own portion of this space. The X.509 defects latent in OpenSSL form a
subset $\Bugs_{\mathsf{OpenSSL}} \subseteq \Bugs_{\mathsf{X.509}}$, those in GnuTLS
another, and so on for every implementation. Cast in these terms, the
current approaches above amount to an open-ended attempt to exhaust these
subsets one at a time. For a standardized protocol this scales poorly, since the
standard is implemented not once but dozens of times across languages and
ecosystems, and an effort that exhausted $\Bugs_{\mathsf{OpenSSL}}$ would
leave $\Bugs_{\mathsf{GnuTLS}}$, $\Bugs_{\mathsf{wolfSSL}}$, and other
implementations' subsets untouched. One might instead hope to search all
implementations at once, covering the union
$\bigcup_{\ell \in \mathsf{Libraries}} \Bugs_{\ell}$ over every library in a
single sweep rather than each subset in sequence.

Such a search is available, but only for bugs of a particular shape:
\textbf{1)} the bug must be detectable by a test case, an input on which an
affected implementation produces the wrong output; and \textbf{2)} the test case
must be expressed in a standardized format, so that the same input can be
submitted to every library alike. For most general-purpose software these
conditions would be a significant restriction, but not necessarily for
cryptographic libraries.

We observe that many cryptographic libraries satisfy the second condition by
construction: they are typically invoked to serve a single, well-defined purpose,
validating a certificate, securing a connection, or checking a token. Since their
responsibilities are so narrow, they expose correspondingly few interfaces, and
those interfaces are often fixed by the specifications they implement. This
restriction also mirrors how these libraries are attacked in practice: downstream
applications consume them through exactly those interfaces, and it is there that
an attacker engages them, by providing a certificate, supplying a JWT token, or
relaying a signed SAML message, each of which the application hands to the
library beneath it for a security decision. To meet the first condition, \sys substitutes
differential testing for the expected-output oracle: when many implementations of
one specification evaluate the same input, agreement stands in for the expected
output, and any disagreement signals that at least one implementation is wrong.

Differential testing over a shared interface amplifies each test case
(Figure~\ref{fig:efficiency-amplified-testing}): one generated input is evaluated
by every harnessed library at once, so a single probe searches not
$\Bugs_{\mathsf{OpenSSL}}$ alone but
$\bigcup_{\ell \in \mathsf{Libraries}} \Bugs_{\ell}$, amortizing its cost across
all libraries. And since cryptographic libraries largely meet the two conditions,
this amplification arrives without the significant restriction it would
impose on other software. Amplification, however, is only as good as the inputs
being amplified. Differential testing has yielded findings at various levels of
significance~\cite{brubakerUsingFrankencertsAutomated2014,
petsiosNEZHAEfficientDomainIndependent2017, chauAnalyzingSemanticCorrectness2019,
parachaHallucinatingCertificatesDifferential2026a}, though current
implementations still carry significant limitations.

The effectiveness of this search therefore rests on the quality of the
generated inputs, since a discrepancy surfaces only when an input reaches a
behavior on which implementations disagree. Grammar-based fuzzers address this
for a single protocol, but each requires a protocol-specific grammar that
encodes the structure and constraints of the message format. For protocols
built on flexible representations such as XML, where the same semantic content
can be expressed in many syntactically distinct ways, these grammars can be
difficult to write and
maintain~\cite{somorovskySystematicFuzzingTesting2016a,mengLargeLanguageModel2024a,maOneThousandPages2024}.
More importantly, the engineering does not generalize: a fuzzer built for SAML
cannot be reused for JWT or digital signature testing without substantial
rework. \sys instead drives the search with agents, which offer two properties
a fixed grammar cannot. First, agents draw on a vast store of protocol
knowledge directly, with no need to distill it into per-protocol grammars or
heuristics, so supporting a new protocol amounts to harnessing its libraries
and supplying seed messages. Second, the search becomes adaptive: an agent can
reason about why the libraries accepted or rejected its last input and steer
the next mutation accordingly, where a conventional fuzzer can only continue
sampling from its fixed distribution (Section~\ref{sec:design}).

\subsection{\sys's Discrepancy Tracing}
\label{sec:tracing}

Prior differential approaches have focused attention on the
libraries: a finding is reported when one implementation deviates
from the specification or from its peers, and the analysis ends there~\cite{parachaHallucinatingCertificatesDifferential2026a,zhuGuidedDeepTesting2020,petsiosNEZHAEfficientDomainIndependent2017,brubakerUsingFrankencertsAutomated2014}.
Some have begun to examine downstream effects, such as mismatches between what
one component verifies and what another consumes~\cite{youMyZIPIsnt2025}
(\textsection\ref{sec:related}). These efforts, however, target concrete
processing mismatches, where two components disagree over the same input within
one system. The discrepancies we pursue are of a different character:
ambiguities, often intentional design decisions that remain defensible at the
specification level, whose hazard lies not in either implementation but in the
security contract assumed between a library and the applications built on it
(Section~\ref{sec:threat-model}). For example, whether a verification key
embedded in the token itself may be trusted: the specification permits it, some
libraries accept such keys, others reject them, and an application that
misreads successful verification as trusted verification is exposed. While
these particular maintainers ultimately made the behavior opt-in
(CVE-2018-0114), other maintainers may not consider the change necessary, and
remain within the specification. We consider that
applications may not always scrutinize this contract when adopting a new
library: the guarantees it does and does not provide can be undocumented or
difficult to communicate.

\sys treats these ambiguities as leads for vulnerabilities in downstream
applications. Each one names a concrete condition, a
library that does not enforce a property its peers do, so the applications at
risk are exactly those that depend on the disagreeing library. Once a specific
ambiguity is identified in a library, discovering potentially insecure
applications is trivial: \sys
queries a dependency database built from package manifests that maps any
library to the projects that
transitively depend on it, and this reverse search
(Figure~\ref{fig:efficiency-reverse-search}) turns the ambiguity into a short
list of candidate applications rather than a sweep over the ecosystem. A single
ambiguity is thus amplified a second time: one library-level discovery extends
to every application the search reaches.

Each candidate then receives a targeted audit
(Figure~\ref{fig:efficiency-targeted-audit}). Because the ambiguity names the
exact behavior in question, the audit reduces to a close-ended check, whether
the application relies on the property its library leaves unenforced, here
whether it trusts a key supplied in the token, and an agent can settle it with
a concrete proof of concept where the application is affected. This stands in
contrast to an open-ended search of the application's entire bug space
(Figure~\ref{fig:audit-contrast}), where the agent must both guess what to look
for and substantiate whatever it finds.
Section~\ref{sec:design} details the dependency graph construction and the
audit procedure.
 \section{System Design}
\label{sec:design}

Differences between a specification and its implementations are inevitable,
whether due to bugs, ambiguities, or latitude granted to implementers. \sys
surfaces these with differential testing: where two libraries disagree on the
same input, at least one deviates from the specification's intent. Turning
that principle into a system poses three problems: generating inputs that
drive libraries into disagreement, allocating effort across a search no single
input can cover, and confirming and classifying the
disagreements that result. A final stage then traces each confirmed disagreement
through the stack to the downstream applications that inherit it.

\begin{figure*}[t]
  \centering
  \includegraphics[width=\textwidth]{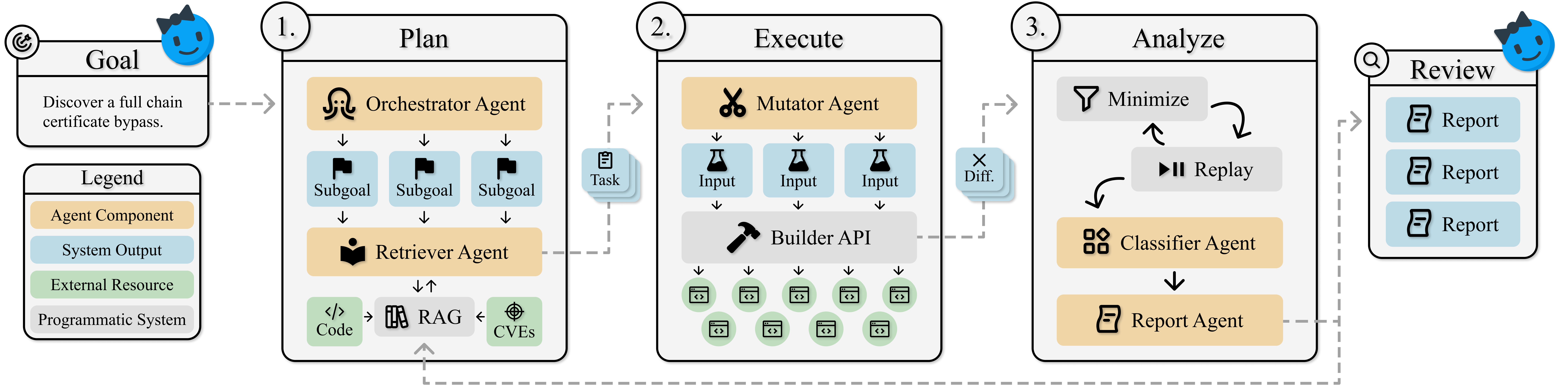}
  \caption{\textbf{Chai Differential Testing.} End-to-end design of \sys. From a high-level goal, \textbf{1)} Plan decomposes it into subgoals with RAG over library source and CVEs, \textbf{2)} Execute mutates and builds candidate inputs for every harness, and \textbf{3)} Analyze minimizes, classifies, and reports each differential. Findings go to review; results feed back into the loop (dashed).}
  \label{fig:design-end-to-end}
\end{figure*}
 
\subsection{Differential Testing: Mutation, Construction, and Adaptive Iteration}
\sys discovers differentials through an iterative loop: a mutation agent
proposes a change, the builder realizes it into a signed message, and every harness
evaluates the identical bytes. Throughout the iterations, each disagreement
plus a memory of prior attempts steers the next proposal. Three concerns structure the loop:
\emph{mutation} decides what to try, \emph{construction} turns it into
reproducible bytes, and \emph{adaptive iteration} steers the search toward the
regions that produce differentials. We take each in turn.

\noindent\textbf{Mutation.} Each search begins from a seed, a known-good baseline
the builder generates from semantic parameters. \sys submits this baseline
unmodified to confirm the libraries agree before mutation begins. From that seed,
the mutation agent proposes a mutation program, an ordered list of operations
each drawn from a fixed catalog that spans the protocol's trust boundaries, with
their arguments left to the agent. Together, the seed and its mutation program
can form a candidate when given to the builder.

\noindent\textbf{Construction.} The builder, not the agent, deterministically
assembles each candidate from a (seed, mutation program) pair. Producing a valid
message involves mechanical work such as serialization, encoding, and
cryptographic operations. Leaving it to the builder lets the agent reason the
way a human vulnerability researcher
does~\cite{singerIncalmoAutonomousRedTeaming2026}, about what behavior to probe
rather than how to encode it. The builder is also aware of the protocol's
structure, applying operations relative to protocol boundaries such as
signatures and encryption. Since construction is deterministic, the same seed
and program always yield the same candidate, so any differential it produces
can be replayed and minimized.

\noindent\textbf{Adaptive iteration.} What the agent proposes next is conditioned on
everything the campaign has seen. Before each iteration, a retrieval agent queries an
index of prior vulnerability disclosures, curated from the harnessed libraries' past
CVEs, for patterns relevant to the current target. Promising but shallow differentials
are escalated automatically: an
accept/reject split, for instance, seeds follow-up candidates that try to turn
it into a higher-impact semantic difference. Leases coordinate this search
across parallel agents and suppress duplicated effort, with allocation detailed
below. This grounding is what separates \sys from a fixed-grammar fuzzer or a
single-pass agentic audit. Rather than mutate blindly or scan once, it steers
each step toward the regions that have already produced differentials.

\subsection{Differential Testing: Resource Allocation}
\label{sec:allocation}
To achieve breadth, \sys deploys many searches in
parallel~\cite{mengLargeLanguageModel2024a,jiFirmAgentLeveragingFuzzing2026}.
Raw parallelism, however, delivers overlap rather than coverage. The searches
are steered by the same model and primed by the same index of prior
vulnerability disclosures. The protocol's most publicized attacks dominate
both, so independent proposals converge on the same few
patterns. They linger wherever differentials come
easily~\cite{chenELFuzzEfficientInput2025}, and they re-derive one another's
findings. \sys therefore mediates the searches through a resource allocation
system. It partitions the search space into leasable units, and each search
must hold a lease before proposing. A bandit algorithm reallocates the
leases, crediting a unit only for the new behaviors it produces; a repeated
result counts for nothing.

\noindent\textbf{Mutation groups.} The leasable unit is the mutation group, a
(cluster, operation) pair. Each operation in the catalog targets a specific
validation behavior: tampering with the digest a signature covers exercises
signature integrity, reordering assertions exercises assertion selection,
and injecting a document type declaration exercises the parser. Clusters are
derived from the disclosure index, where each disclosure is tagged with the
attack families it exercised. The families overlap too heavily to partition
the search space.

For instance, most disclosures in our SAML index carry several. As a
result, \sys collapses these attack families into a small set of disjoint
clusters, one per trust boundary. For SAML, these cover signature and key
trust, assertion selection and wrapping, parsing, metadata trust, and
transport encoding. Within a group, the two halves do different work. The
cluster fixes the probe's intent, naming the behavior the mutation tries to
subvert, and steers the retrieval agent toward disclosures of that kind. The
operation fixes the mechanism, the concrete change applied to the candidate.
The distinction is why the same operation may appear under several clusters:
editing an assertion to bypass a signature and editing one to confuse the
parser share a mechanism but not a hypothesis, and \sys leases them as
distinct groups. Neither half alone could serve as the unit. Clusters are
too coarse to allocate over, and bare operations carry no intent.

\noindent\textbf{Lease allocation.} Each search leases one unit from a shared
store, works within it for a bounded set of attempts, and releases the lease
with a record of what the unit produced. To decide which unit a lease
grants, \sys adopts UCB1~\cite{auerFinitetimeAnalysisMultiarmed2002}, an
upper-confidence-bound policy for the multi-armed bandit problem, treating
every unit as an arm and every lease as a pull. Each unit is leased once
before any is leased twice. The campaign therefore sweeps the full space
before exploitation begins, and an early finding cannot capture the budget.
After the sweep, a unit's score is its observed yield plus a bonus that
grows the longer the unit goes untried. The highest-scoring unit is leased
next. Under UCB1, productive units are revisited, and neglected units force
their way back into contention. Units already under lease are passed over,
steering concurrent searches apart.

\subsection{Differential Testing: Output Analysis}

Output analysis turns the search's raw differentials into findings. A
differential is a disagreement among libraries, and only that: it tells us
nothing about whether the disagreement reproduces, which of its mutations are
essential to it, or what kind of issue it represents. The differential may be
a nondeterministic fluke, may carry mutations irrelevant to the divergence,
and may have no security consequence at all. Output analysis resolves these
in turn, producing a replayed, minimized, and classified finding. This is
where \sys delivers the verifiable signal it set out to provide: a
deterministic check that stands in for the expected-output oracle a single
implementation cannot give.

\begin{figure}[t]
  \centering
  \begin{tikzpicture}[
      font=\footnotesize,
      every node/.style={align=left},
      tag/.style={
        font=\footnotesize\bfseries,
        text=black,
        anchor=west,
      },
      promptbox/.style={
        draw=black!45,
        line width=0.4pt,
        rounded corners=2.5pt,
        fill=white,
        text width=64mm,
        inner xsep=7pt,
        inner ysep=5pt,
        anchor=north west,
        fig shadow,
      },
      targetbox/.style={
        promptbox,
        draw=stargold!75!black,
        fill=stargold!7,
      },
    ]

\begin{scope}[shift={(0.11,0.02)}]
      \draw[line width=0.55pt, black!55] (0,0) circle (0.075);
      \draw[line width=0.9pt, black!55, line cap=round]
      (0.053,-0.053) -- (0.10,-0.10);
    \end{scope}
    \node[tag] (lab1) at (0.30,0) {Open-ended Audit};
    \node[promptbox] (open) at (0,-0.26)
    {Try to identify ways for authentication bypass in this application.};

\begin{scope}[shift={(0.11,-1.48)}]
      \draw[line width=0.55pt, stargold!75!black] (0,0) circle (0.085);
      \fill[stargold!75!black] (0,0) circle (0.024);
      \foreach \a in {0,90,180,270}
      \draw[line width=0.55pt, stargold!75!black, line cap=round]
      (\a:0.085) -- (\a:0.135);
    \end{scope}
    \node[tag] (lab2) at (0.30,-1.5) {Targeted Audit};
    \node[targetbox] (target) at (0,-1.76)
    {This project's JWT library ignores non-base64 characters instead of
      rejecting them, so distinct strings yield the same token. Could that
    cause a security problem?};

\begin{scope}[on background layer]
      \node[
        draw=black!55,
        line width=0.5pt,
        rounded corners=4pt,
        fill=black!1.5,
        fit=(lab1)(open)(lab2)(target),
        inner sep=5pt,
      ] {};
    \end{scope}

  \end{tikzpicture}
  \caption{\textbf{Open-ended vs. targeted audits.} Discrepancy tracing yields targeted audits, prompting an agent to check a specific property violation.}
  \label{fig:audit-contrast}
\end{figure}
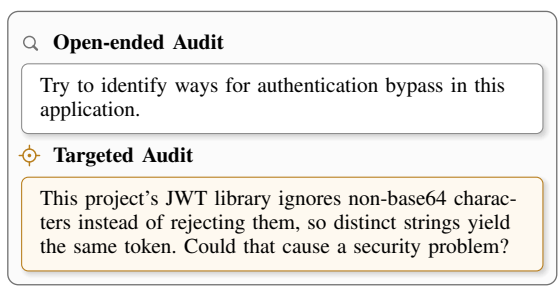
 
\noindent\textbf{Replay and minimization.} Since construction is deterministic, a
candidate's bytes never change between submissions, so a differential that
fails to reproduce exposes nondeterminism in a library rather than in the
input. In addition to ruling out this nondeterminism, \sys isolates what each
differential depends on, looping over the mutation program: it strips one
operation at a time, resubmits, and keeps a removal only when the outcome
bucket is unchanged. The loop ends at the smallest program that still
reproduces the differential.

\noindent\textbf{Classification.} A classifier then labels each differential: what
diverged, and the validation behavior it implicates. It works from the
libraries' normalized verdicts and the mutation that produced them, not the
protocol message, so it organizes the evidence rather than judging validity.
The differential itself remains the signal that something is wrong. These
labels help sort each differential into one of the four threat-model classes
(Section~\ref{sec:threat-model}): vulnerabilities and security bugs go to
reporting, ambiguities into downstream tracing, and inconsequential
differences are dropped (Section~\ref{sec:depgraph}). A person reviews every
resulting report before it leaves \sys.

\subsection{Discrepancy Tracing}
\label{sec:depgraph}

The ambiguities that differential testing surfaces are not yet
  vulnerabilities, but they are security hazards: the risk is that a downstream
  application inherits the weaker
  behavior, having been written against an assumption the library beneath it does
  not enforce. We automate as much of the path from a single ambiguity to a
  finished report and working proof of concept in downstream applications as we
  can. However, we do keep a human in the loop at the end to verify each report
  before filing. We consider this final verification mandatory on ethical grounds,
  so that we do not waste the time of open-source maintainers with AI-generated
  false reports. Much of the stage's design is in service of that review,
  structured as a
  funnel that narrows a broad set of candidate applications down to the few that
matter, so the researcher's attention falls only on substantiated findings.

A benefit of working from a confirmed ambiguity is that it lets us hand each
  agent a narrowly scoped task. Conventional agentic discovery couples two burdens:
  finding which weaknesses might exist and auditing whether any is real. Starting
  from a confirmed ambiguity removes the discovery burden and leaves only the audit
  of whether a single property holds for a given application
  (Figure~\ref{fig:audit-contrast}). Prior work has observed
  that agents perform more reliably on shorter, well-scoped tasks than on broad
open-ended ones~\cite{singerIncalmoAutonomousRedTeaming2026}.

\noindent\textbf{Stage 1: Enumeration.} To enumerate the applications a given ambiguity
  can reach, we construct our own dependency graph from package manifests across
  the open-source ecosystem. We seed it from the 50{,}000 most critical
  repositories by OpenSSF Criticality Score~\cite{openssfCriticalityScore},
  scanning each for manifests across 17 ecosystems including npm, PyPI, Go modules,
  Maven, Cargo, and Composer. From every manifest we record which packages depend
  on which and follow those links, including indirect dependencies, so that an
  ambiguity in a library resolves to the projects that inherit its behavior even
through intermediate packages.

\noindent\textbf{Stage 2: Initial pass.} A coding agent then audits each candidate one
  at a time, beginning with a lightweight read-only pass over its configuration and
  use of the library to judge whether the ambiguity could plausibly be exposed, for
  example whether the application trusts a key supplied in the token. This yields a
  shortlist of susceptible applications; for each, the agent attempts a proof of
  concept in a sandboxed virtual machine, and those whose attempt produces a
genuine programmatic result are promoted to the next stage.

\noindent\textbf{Stage 3: Proof-of-concept pipeline.} This stage takes a promoted
  application and builds a full proof of concept (PoC). PoCs land on a spectrum in
  terms of what they prove, from a unit test that isolates the bug in an internal
  function to an end-to-end exploit driven through the deployed application; we aim
  for the latter to prioritize the findings with the clearest impact. We task
  coding agents to generate several candidate PoCs independently, each on its own
  disposable spot VM as a sandbox that stands up the whole application, often
  several Docker containers, to improve reproducibility. The pipeline first
  iterates to push each PoC as far
  toward the end-to-end end of the spectrum as it can, then minimizes its changes
  to the application's default configuration to remove unnecessary assumptions,
  with a judge picking a winner each round. The final harness is re-validated on a
  clean VM and delivered as three files: a setup script, a run script that records
evidence, and a unified git diff of all supporting artifacts.

\noindent\textbf{Stage 4: Report generation.} The final stage assembles a disclosure
  around the validated PoC: it structures the finding, documents the proof
  artifacts, and runs a prior-art search over public issue trackers and disclosed
  CVEs to flag duplicates and to better gauge precedent and severity, before
  rendering drafts tailored both to the maintainer disclosure and to our own
  internal security review. Reports produced here are always manually reviewed, and
  their text revised, prior to disclosure. We reuse this same stage for
  the library-level findings from the first stage, which are already concrete once
  reviewed and so bypass enumeration and the proof-of-concept pipeline; reproducing
  them rarely requires standing up an application stack, and typically amounts to a
short script against the installed library.

 \section{Implementation}
\label{sec:implementation}

\sys spans three protocol domains, each a complete differential testing
  system over a family of real implementations. For certificates, it harnesses 13
  X.509 path-validation libraries, including OpenSSL, BoringSSL, and GnuTLS. For
  JWT and JOSE, it exposes 23 runnable harnesses. For SAML, it harnesses 11
  service-provider libraries. The X.509, JWT, and SAML systems comprise roughly
  20k, 15k, and 17k lines of code, with a further 11k for the report pipeline and
  web visualizer, and 1.4k for the proof-of-concept controller.

The X.509, JWT, and SAML differential testing systems are each written in
  Python. The builder API uses standard cryptographic and parsing libraries:
  \texttt{cryptography}, \texttt{asn1crypto}, and \texttt{pyOpenSSL} for X.509,
  \texttt{cryptography} and \texttt{ecdsa} for JWT, and \texttt{signxml} and
  \texttt{lxml} for SAML. Each candidate is encoded in its protocol's wire format,
  a DER-encoded certificate chain for X.509, a compact JWS or JWE token for JWT,
  and a signed XML response for SAML, before submission to the harnesses. Agent
  requests route through LiteLLM to track spend and route to multiple providers,
  with non-LLM fuzzers available as baselines. For the naive agent baselines, we run
Codex CLI 0.137.0 and Claude Code 2.1.173.

Each harness wraps a library behind a uniform interface and is implemented as
  a small native-language script invoked as a subprocess, exchanging a serialized
  message and a structured JSON verdict over \texttt{stdin} and \texttt{stdout}.
  The verdict records the accept or reject decision, any rejection reason or
  exception, and any identity attributes the library extracts. Libraries written in
  C, Go, Ruby, PHP, and Node.js are harnessed directly; the native X.509 libraries
  are reached through thin Python wrappers over their bindings. Harnesses are
  authored with the assistance of a coding agent and validated against a shared
suite of known-good and known-bad inputs for the protocol.

The proof-of-concept controller is a typed Python CLI that enforces strict
  types on the shape of each proof of concept and uses the GCP SDK to launch
  validation VMs in parallel. The reporting stage renders JSON and Markdown reports
  through Jinja templates and searches for related GitHub issues and CVEs from the
  NVD database. A separate web visualizer, built in React and backed by Postgres,
renders evidence and report text with Markdown and diff views for review.

\begin{table}[t]
  \centering
  \caption{Zero-day vulnerabilities and security bugs discovered. Details are
  intentionally redacted while disclosures are in progress.}
  \label{tab:protocol-layer-findings}
  \footnotesize
  \setlength{\tabcolsep}{1pt}
  \begin{tabular}{@{}p{58pt}>{\raggedright\arraybackslash}p{123pt}rr@{}}
\toprule
Library & Description & Stars & LLoC \\
\midrule
\multicolumn{4}{l}{\textbf{X.509}} \\
\midrule
wolfSSL & {\tiny\textbullet}\hspace{0.35em}Chain validation bypass (SKI) & $>$2k & $>$2M \\
 & {\tiny\textbullet}\hspace{0.35em}Chain validation bypass (depth) &  &  \\
X.509 Library A & {\tiny\textbullet}\hspace{0.35em}EKU/KeyUsage fail-open & $>$100 & $>$1M \\
X.509 Library B & {\tiny\textbullet}\hspace{0.35em}Unenforced intermediate constraints & $>$100 & $>$200k \\
X.509 Library C & {\tiny\textbullet}\hspace{0.35em}Host checking issue & $>$200 & $>$20k \\
\midrule
\multicolumn{4}{l}{\textbf{JWT/JOSE}} \\
\midrule
JWT Library A & {\tiny\textbullet}\hspace{0.35em}Fail-open malformed header & $>$5k & $>$20k \\
JWT Library B & {\tiny\textbullet}\hspace{0.35em}Fail-open malformed header & $>$200 & $>$5k \\
JWT Library C & {\tiny\textbullet}\hspace{0.35em}Verification-policy flaw & $>$5k & $>$2k \\
JWT Library D & {\tiny\textbullet}\hspace{0.35em}Verification-policy flaw & $>$5k & $>$5k \\
 & {\tiny\textbullet}\hspace{0.35em}Malformed-claim acceptance &  &  \\
 & {\tiny\textbullet}\hspace{0.35em}Claim-validation bypass &  &  \\
JWT Library E & {\tiny\textbullet}\hspace{0.35em}Algorithm confusion & $>$2k & $>$100k \\
JWT Library F & {\tiny\textbullet}\hspace{0.35em}Algorithm confusion & $>$5k & $>$5k \\
JWT Library G & {\tiny\textbullet}\hspace{0.35em}Verification-policy flaw & $>$2k & $>$5k \\
JWT Library H & {\tiny\textbullet}\hspace{0.35em}Verification-policy flaw & $>$2k & $>$50k \\
\midrule
\multicolumn{4}{l}{\textbf{SAML}} \\
\midrule
SAML Library A & {\tiny\textbullet}\hspace{0.35em}Signed-attribute parsing flaw & $>$1k & $>$100k \\
\bottomrule
\end{tabular}
 \end{table}
 \section{Evaluation}
\label{sec:evaluation}

\begin{figure*}[t]
  \centering
  \captionsetup[subfloat]{labelfont=rm,textfont=rm}
  \begin{minipage}[c]{0.79\textwidth}
    \centering
    \begin{minipage}[c]{0.327\linewidth}
      \centering
      \subfloat[X.509]{\includegraphics[width=\linewidth]{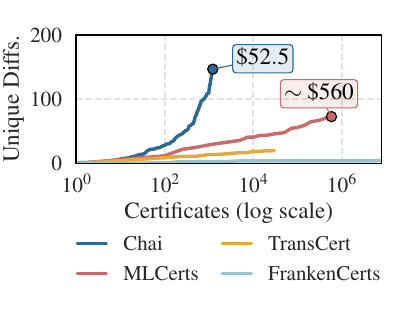}\label{fig:findings-over-executions-ai-text}}
    \end{minipage}\hfill
    \begin{minipage}[c]{0.327\linewidth}
      \centering
      \subfloat[JWT]{\includegraphics[width=\linewidth]{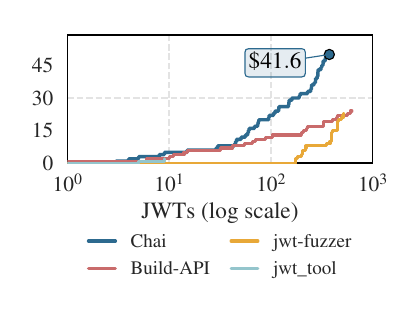}\label{fig:findings-over-executions-b}}
    \end{minipage}\hfill
    \begin{minipage}[c]{0.327\linewidth}
      \centering
      \subfloat[SAML]{\includegraphics[width=\linewidth]{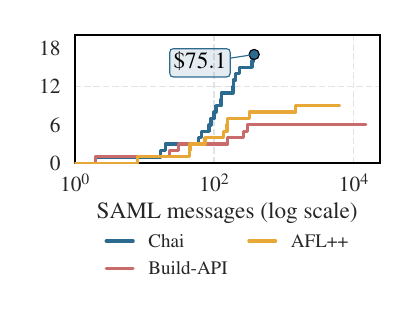}\label{fig:findings-over-executions-c}}
    \end{minipage}
  \end{minipage}\hfill
  \begin{minipage}[c]{0.18\textwidth}
    \caption{Unique differences surfaced by \sys and baselines over 1,500 executions, where one execution is a single input (X.509 certificate, JWT, or SAML message) submitted to all harnessed libraries. Annotations give total cost, included only for methods that incur inference or GPU cost.}
    \label{fig:findings-over-executions}
  \end{minipage}
\end{figure*}
 
\subsection{Experimental Setup}
\label{sec:disclosures}

The differential testing stage runs against the harnessed X.509, JWT, and
SAML libraries described in Section~\ref{sec:implementation}, seeded with a corpus of high-severity CVEs
drawn from each domain. For the discrepancy tracing stage, we build the
dependency graph from the latest OpenSSF Criticality Score (2025.07.25).
We evaluate \sys across four models: GPT-5.5
(\texttt{gpt-5.5-2026-04-23}), Gemini 3.5 Flash, and Claude Opus 4.8, along with
one open-source model, Kimi K2.6, for reproducibility. Each runs at its
provider's default temperature, accessed through the OpenAI API and OpenRouter. Retrieval-augmented steps use
OpenAI's \texttt{text-embedding-3-small} embedding model with cosine
similarity.

\begin{figure*}[t]
  \centering
  \captionsetup[subfloat]{labelfont=rm,textfont=rm}
  \begin{minipage}[c]{0.72\textwidth}
    \centering
    \begin{minipage}[c]{0.32\linewidth}
      \centering
      \subfloat[X.509]{\includegraphics[width=\linewidth]{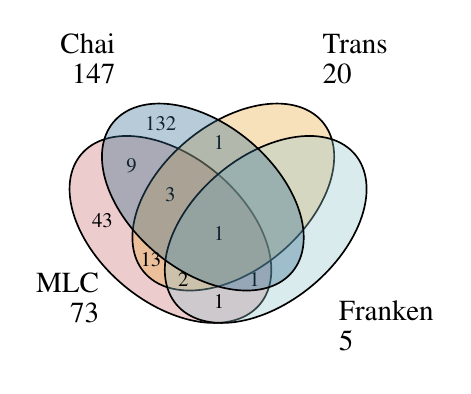}\label{fig:baseline-overlap-grid-x509}}
    \end{minipage}\hfill
    \begin{minipage}[c]{0.32\linewidth}
      \centering
      \subfloat[JWT]{\includegraphics[width=\linewidth]{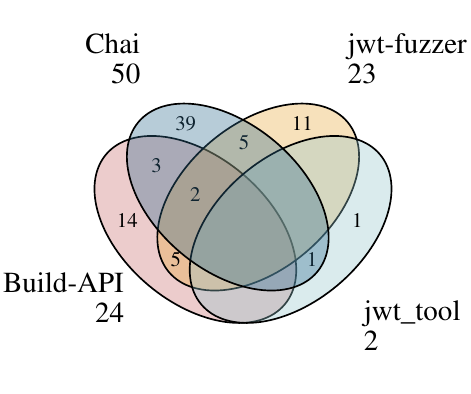}\label{fig:baseline-overlap-grid-jwt}}
    \end{minipage}\hfill
    \begin{minipage}[c]{0.32\linewidth}
      \centering
      \subfloat[SAML]{\includegraphics[width=\linewidth]{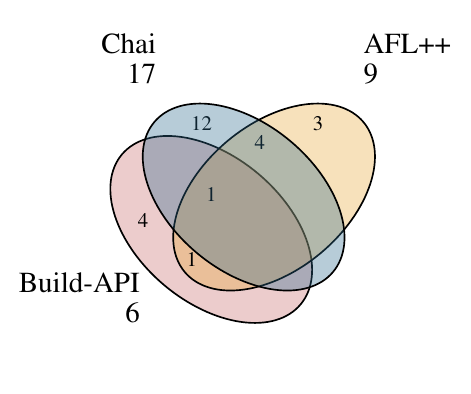}\label{fig:baseline-overlap-grid-saml}}
    \end{minipage}
  \end{minipage}\hfill
  \begin{minipage}[c]{0.25\textwidth}
    \caption{Overlap of unique differences found by \sys and baselines, per
        protocol, over the same harnessed libraries. In each protocol, the large majority of differences
        \sys surfaces are found by no baseline. Differences are
        deduplicated by the per-protocol equivalence notion of
        Section~\ref{sec:evaluation}.}
    \label{fig:baseline-overlap-grid}
  \end{minipage}
\end{figure*}
 \begin{figure*}[t]
  \centering
  \captionsetup[subfloat]{labelfont=normalfont,textfont=normalfont}
  \subfloat[\normalfont X.509]{
    \includegraphics[width=0.32\textwidth]{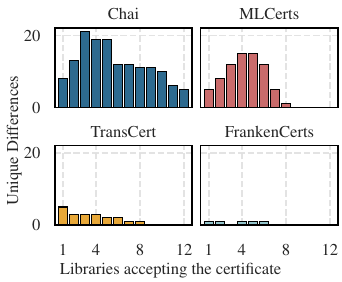}\label{fig:accept-count-distributions-x509}}\hfill
  \subfloat[\normalfont JWT]{
    \includegraphics[width=0.32\textwidth]{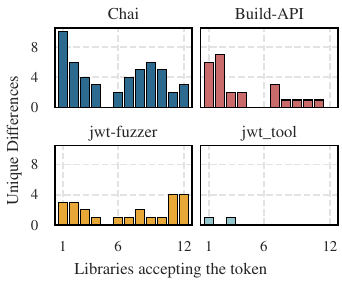}\label{fig:accept-count-distributions-jwt}}\hfill
  \subfloat[\normalfont SAML]{
    \includegraphics[width=0.32\textwidth]{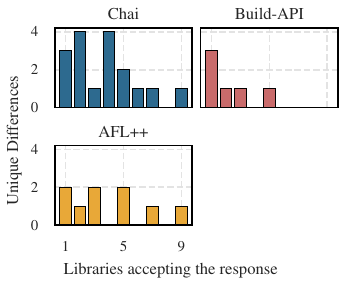}\label{fig:accept-count-distributions-saml}}
  \caption{Distribution of accept counts across X.509, JWT, and SAML test inputs.}
  \label{fig:accept-count-distributions}
\end{figure*}

We first measure unique differences as prior differential-testing work does, by reducing each library's outcome to a success or failure binary vector~\cite{parachaHallucinatingCertificatesDifferential2026a}. For SAML and JWT, due to their lower complexity, we normalize errors to gain further granularity in discrepancy counts. These unique differences thus provide insight into the breadth of discoveries found by \sys. When comparing against related work, we account
  for inference and GPU costs but not CPU-bound compute such as fuzzing and
  differential testing, in line with prior LLM-driven security
evaluations~\cite{fangLLMAgentsCan2024a}. For protocols with minimal baselines (JWT, SAML), we evaluate \sys against a naive fuzzer constructed to send inputs via the Build-API, which removes the agentic reasoning component from \sys's discovery, and additionally evaluate SAML against an AFL++~\cite{fioraldiAFLCombiningIncremental2020} baseline with an XML dictionary.

\noindent\textbf{Vulnerability Disclosure Process:} All vulnerabilities reported in Table~\ref{tab:protocol-layer-findings} were confirmed by the authors, and all X.509 vulnerabilities have been disclosed and acknowledged by maintainers. Validation
  is deliberate and labor-intensive: each disclosure requires roughly 2 to 5 hours to
  prepare, as we work through the target project's context, learn the relevant
  library details, and ensure we are not submitting false reports that would
  impose undue burden on maintainers. As such, disclosure is ongoing. We prioritize by severity, disclosing
  library-level findings before downstream-application findings, since a single
  library vulnerability can affect thousands of dependent projects. The most severe
  findings were disclosed within 24 hours of initial discovery to minimize the
window in which they could be exploited.

\subsection{Evaluation Summary}
\label{sec:eval-summary}

\sys surfaced over 100 vulnerabilities and security bugs across X.509, JWT, and SAML
(Table~\ref{tab:protocol-layer-findings}). The most severe affect libraries in
near-ubiquitous use. In X.509 alone, \sys found a critical chain-validation
bypass in wolfSSL, whose TLS stack ships in billions of devices; a certificate-constraint fail-open in the TLS library behind a major browser; and a certificate-chain validation flaw in a TLS library shipped in major Linux distributions. Every finding reported here
was confirmed by the authors and is being disclosed to maintainers
(Section~\ref{sec:disclosures}).

\sys's findings come from both of its stages: differential search at the
library level, and discrepancy tracing in the applications above. Across all
three protocols, the library findings share a single failure mode: a library
accepts input it should reject, whether an invalid certificate chain, a forged
token, or a tampered assertion (Table~\ref{tab:protocol-layer-findings}). The
downstream findings come from tracing library-level ambiguities into the
applications that inherit them, and make up the remainder.

The rest of this section evaluates these results against the alternatives and
dissects the design behind them. Section~\ref{sec:related-eval} compares \sys's
differential testing against prior differential and fuzzing tools, in both the
rate at which it surfaces unique disagreements and the kinds it surfaces.
Section~\ref{sec:naive-agent} contrasts \sys with general-purpose coding
agents, Claude Code and Codex, examining the same library, isolating the
value of starting from a concrete discrepancy rather than searching
open-endedly. Section~\ref{sec:design-eval} examines \sys's own design,
measuring cost and yield across models and identifying its most productive
mutation groups. Section~\ref{sec:downstream} follows discrepancies into
downstream applications, and Section~\ref{sec:disclosures} reports the
disclosure status of each finding.

\subsection{Comparison to Existing Tools}
\label{sec:related-eval}

We first ask whether \sys's adaptive search surfaces differentials more
efficiently than prior differential testing and fuzzing.
Figure~\ref{fig:findings-over-executions} plots the unique differentials each
tool finds against the inputs it submits, across X.509, JWT, and SAML. On
every protocol \sys's curve rises faster and reaches higher, finding more
unique differentials from fewer inputs. The contrast is sharpest on X.509:
\sys surfaces 147 unique discrepancy vectors from 1,500
certificates at \$52.5, where the strongest baseline, MLCerts, reaches 73 only
after ${\sim}500,000$ certificates at ${\sim}$\$560: twice as many differentials
at a tenth the cost and a thousandth the inputs. \sys leads on JWT and SAML as
well. On JWT, jwt-fuzzer and jwt\_tool run to exhaustion yet still surface
fewer unique differentials than \sys, whose curve has not yet plateaued.

\begin{figure*}[t]
  \centering
  \makebox[\textwidth][c]{\includegraphics[width=.339\textwidth]{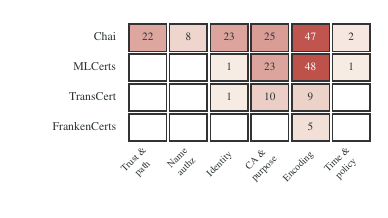}\hspace{-.012\textwidth}\includegraphics[width=.339\textwidth]{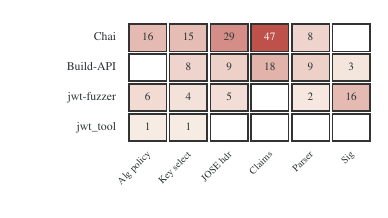}\hspace{-.012\textwidth}\includegraphics[width=.339\textwidth]{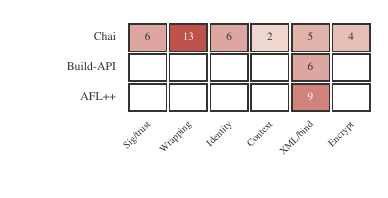}}\\[-.2em]
  \makebox[\textwidth][c]{\makebox[.339\textwidth][c]{\normalfont\footnotesize (a) X.509}\hspace{-.012\textwidth}\makebox[.339\textwidth][c]{\normalfont\footnotesize (b) JWT}\hspace{-.012\textwidth}\makebox[.339\textwidth][c]{\normalfont\footnotesize (c) SAML}}
  \caption{Unique differences found by \sys and baselines per protocol, grouped by the validation behavior each difference exercises. Blank cells indicate categories where a tool surfaced no differences. A higher-fidelity breakdown of each category appears in Appendix~\ref{app:detailed-capability-heatmaps}.}
  \label{fig:capability-heatmaps-condensed}
\end{figure*}
 
\begin{table*}[t]
  \centering
  \caption{Blind wolfSSL rediscovery runs for Claude Code and Codex.}
  \label{tab:wolfssl-subscription-runs}
  \setlength{\tabcolsep}{3.5pt}
  \begin{tabularx}{\textwidth}{@{}p{34pt}p{62pt}p{48pt}p{48pt}XXXp{30pt}p{24pt}@{}}
\toprule
Hint & System & Model & Time (min) & Input tokens & Output tokens & Cost (USD) & Depth & SKI \\
\midrule
\multirow{2}{*}{{\footnotesize None}} & {\footnotesize Claude Code} & {\footnotesize Opus 4.8} & {\footnotesize 8.1 $\pm$ 3.0} & {\footnotesize 1.88M $\pm$ 919.1k} & {\footnotesize 31.0k $\pm$ 9.7k} & {\footnotesize \$2.15 $\pm$ \$0.79} & {\footnotesize 10/10} & {\footnotesize 0/10} \\
 & {\footnotesize Codex} & {\footnotesize GPT-5.5} & {\footnotesize 8.8 $\pm$ 1.8} & {\footnotesize 4.02M $\pm$ 974.7k} & {\footnotesize 21.9k $\pm$ 4.3k} & {\footnotesize \$3.83 $\pm$ \$1.05} & {\footnotesize 0/10} & {\footnotesize 0/10} \\
\addlinespace[0.5em]
\multirow{2}{*}{{\footnotesize File}} & {\footnotesize Claude Code} & {\footnotesize Opus 4.8} & {\footnotesize 18.7 $\pm$ 7.1} & {\footnotesize 6.40M $\pm$ 2.66M} & {\footnotesize 82.8k $\pm$ 28.3k} & {\footnotesize \$6.22 $\pm$ \$2.37} & {\footnotesize 7/10} & {\footnotesize 1/10} \\
 & {\footnotesize Codex} & {\footnotesize GPT-5.5} & {\footnotesize 8.1 $\pm$ 2.0} & {\footnotesize 2.57M $\pm$ 965.3k} & {\footnotesize 18.1k $\pm$ 3.5k} & {\footnotesize \$2.60 $\pm$ \$0.75} & {\footnotesize 0/10} & {\footnotesize 0/10} \\
\addlinespace[0.5em]
\multirow{2}{*}{{\footnotesize Commit}} & {\footnotesize Claude Code} & {\footnotesize Opus 4.8} & {\footnotesize 15.2 $\pm$ 5.9} & {\footnotesize 3.81M $\pm$ 2.34M} & {\footnotesize 56.3k $\pm$ 24.1k} & {\footnotesize \$3.90 $\pm$ \$1.98} & {\footnotesize 0/10} & {\footnotesize 0/10} \\
 & {\footnotesize Codex} & {\footnotesize GPT-5.5} & {\footnotesize 13.2 $\pm$ 1.9} & {\footnotesize 3.28M $\pm$ 699.5k} & {\footnotesize 21.4k $\pm$ 2.3k} & {\footnotesize \$3.22 $\pm$ \$0.56} & {\footnotesize 0/10} & {\footnotesize 0/10} \\
\bottomrule
\end{tabularx}
 \end{table*}

We next investigate whether \sys finds the same differentials as prior work. Figure~\ref{fig:baseline-overlap-grid} partitions the unique
accept/reject vectors by which tools surfaced them, with a Venn diagram for
each protocol. On all three, the large majority of \sys's findings are \sys's
alone: 132 of 147 on X.509, 39 of 50 on JWT, and 12 of 17 on SAML. \sys is not
a faster route to the same findings but a route to different ones. The overlap
runs both ways: each baseline surfaces some vectors \sys misses, such as the 43
unique to MLCerts on X.509, so the approaches are partly complementary. Even
so, \sys's unique set is the largest on every protocol, roughly three times the
size of any baseline's.

\sys also finds a more useful kind of differential.
Figure~\ref{fig:accept-count-distributions} bins each unique difference by how
many of a protocol's libraries accept the input, and on all three protocols
\sys's distribution differs from the baselines in two ways. It is broader:
where MLCerts clusters in a tight bell around even splits, surfacing almost
nothing accepted by more than eight of the twelve X.509 libraries, and the JWT
and SAML fuzzers occupy similarly narrow bands, \sys spreads across the entire
range. And it leans left, placing much of its mass where only a few libraries
accept an input the rest reject. That left tail is the hardest region to reach
and the most valuable: a small accepting minority against a rejecting majority
pinpoints the deviation to those few libraries, a strong signal of a
library-level vulnerability, whereas an even split leaves which behavior is
correct unresolved. Digging further, Figure \ref{fig:capability-heatmaps-condensed} depicts how \sys captures a wide breadth of differentials that the existing approaches do not.

\subsection{Comparison to Open-Ended Discovery}
\label{sec:naive-agent}

A prevailing paradigm for agentic vulnerability discovery is open-ended:
point an agent at a codebase and ask it to find flaws. \sys instead starts
from a concrete differential and confirms it. To isolate what that inversion
buys, we test the open-ended approach on the two wolfSSL certificate bypass vulnerabilities. We take a single library,
wolfSSL, that \sys already showed to contain two chain-validation bypasses, and
ask two frontier agents, Claude Code (Opus 4.8) and Codex (GPT-5.5), to find a
certificate-validation flaw in it, ten runs each, with and without a hint
naming the vulnerable file (Table~\ref{tab:wolfssl-subscription-runs}).
Even with advantages, rediscovery is unreliable. Codex finds neither
bypass in any of its ten runs. Claude Code finds the depth bypass readily,
seven of ten unprompted, yet finds the subtler SKI bypass only once across all of
its runs. 

\sys finds the same flaws without auditing the source. Rather than inspect
wolfSSL's implementation for a suspected bug, it searches for inputs on which
wolfSSL disagrees with its peers, and the disagreement localizes the flawed
behavior from the libraries' outputs alone. Both the SKI and depth bypasses surface this way,
among the 147 X.509 differentials \sys found for \$1.07 and 2.05, respectively. The model still
proposes the inputs, but a concrete, reproducible discrepancy replaces the
code-level intuition the open-ended agents could not reliably summon.

\subsection{Chai Design Evaluation}
\label{sec:design-eval}

\begin{figure}[t]
  \centering
  \includegraphics[width=\linewidth]{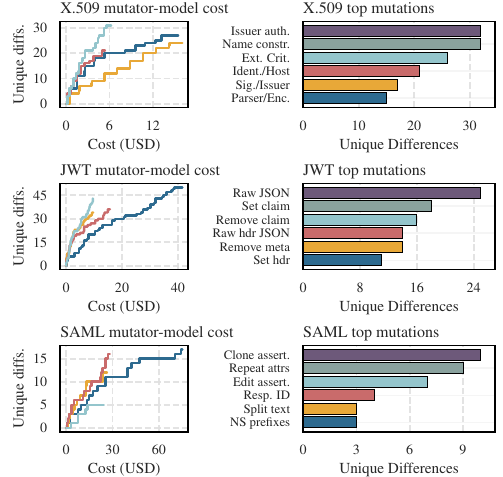}
  \includegraphics[width=.98\linewidth]{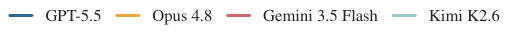}
  \caption{Per-protocol breakdown of \sys's differential search. Left: cumulative unique differences against inference cost (USD) with each of the four mutator models of Section~\ref{sec:evaluation}. Right: the mutation groups (Section~\ref{sec:design}) that produced the most unique differences.}
  \label{fig:strata-efficiency}
\end{figure}

\sys's discovery loop is agent-driven, yet its effectiveness does not rest on
any single model. Figure~\ref{fig:strata-efficiency} runs the same X.509, JWT, and
SAML campaigns under four models, three proprietary (GPT-5.5, Opus 4.8,
Gemini 3.5 Flash) and one open-source (Kimi K2.6), plotting unique
differentials against dollar cost. All four find differentials on every
protocol for tens of dollars, and no single model dominates: the leader shifts
from protocol to protocol. The pipeline runs on the open-source model as well,
so \sys's advantage rests on its design rather than on proprietary frontier
access, and should only grow more scalable as capable models become cheaper
and more accessible.

Figure~\ref{fig:strata-efficiency} also attributes each protocol's differentials to
the mutation group that produced them, and the yield is markedly uneven. A
handful of groups are far more rewarding than the rest, and which ones they are
is protocol-specific. On X.509, issuer-authority and name-constraints mutations
each surface roughly twice what parser and encoding mutations do. On JWT,
raw-JSON manipulation leads by a wide margin, and it is the group that surfaced
a JWT audience-confusion bypass in one library. \sys's lease allocation
(Section~\ref{sec:allocation}) is built for exactly this terrain: it
concentrates effort on the rewarding groups while still sweeping every one, so
it mines the rich seams without having to know in advance which they are.

\subsection{Downstream Tracing}
\label{sec:downstream}

Discrepancy tracing turns a library-level ambiguity into a downstream search, so its value rests on choosing ambiguities that consuming applications actually inherit. Among those \sys surfaced, the JWT Base64URL-decoding ambiguity and SAML assertion replay (Section~\ref{sec:case-studies}) were the most convincing, and we evaluate tracing on both. 
The choice reflects an interesting insight where we found latitude lives across our three protocols. Indeed, X.509 carries significant internal complexity, evident in both the volume of discrepancies \sys surfaces and the size of the implementations that handle it. That complexity, however, we suggest is more concentrated in its libraries rather than the applications, reaching it through operating-system or runtime interfaces that validate certificates on their behalf. JWT and SAML are simpler protocols, yet they leave far more room for latitude and error in the consuming application, which assembles the verification itself and decides what each accepted token or assertion may do. The downstream hazard therefore concentrates in JWT and SAML, which is where we direct tracing.

We trace each ambiguity through the dependency graph to the projects that consume the disagreeing libraries. The reverse search returns far more candidate dependents than can be audited by hand, so we rank them by OpenSSF criticality score and shortlist the top 300. We are currently auditing and triaging 159 of these projects for which the initial audit passes returned a proof of concept, though we have prioritized disclosing library-level vulnerabilities first, since a single library flaw can affect thousands of dependents.

 \section{Case Studies}
\label{sec:case-studies}

To avoid enabling reconstruction of unresolved vulnerabilities while coordinated
disclosures are ongoing, we redact target names and selected implementation
details for non-public findings. We retain the vulnerability class, validation
method, and security impact.

\noindent\textbf{Vulnerability: wolfSSL Certificate-Chain Bypass.} \sys found a
previously unknown full
certificate-chain validation bypass in
wolfSSL. X.509 chain validation should accept a leaf certificate only if it can be
traced, through one or more intermediate certificates, back to a root the verifier
already trusts in its trust store. Applications that use wolfSSL's
OpenSSL-compatible API run this check via the \texttt{WOLFSSL\_X509\_STORE\_CTX}
interface, handing it the leaf to verify, the set of trusted roots, and the
intermediate certificates the other party supplied. To check the leaf's signature,
wolfSSL temporarily loads each
supplied intermediate into its certificate manager, the internal set of
certificates it treats as trusted issuers, just long enough to confirm that the
intermediate actually signed the leaf. If that
intermediate then fails to connect to a trusted root, wolfSSL should unload it
again and reject the chain. However, under certain edge cases it does not: the intermediate is
left loaded in the certificate manager, so it goes on to vouch for the leaf, and
verification succeeds even though nothing in the chain reaches a trusted root. Given
the same chains, two other TLS libraries reject, while
wolfSSL reports success.

The precise cause is in the retry logic of \texttt{wolfSSL\_X509\_verify\_cert()}
in \texttt{src/x509\_str.c}: the step that unloads the intermediate keys removal
to the certificate's Subject Key Identifier, an optional field that labels a
certificate. An intermediate that omits this field is never matched, so it is never
unloaded and stays behind as a valid issuer for the leaf.

The flaw is reached by more than one broken chain: besides an intermediate
that reaches no trust anchor at all, it also accepts a chain anchored to a trusted
but unrelated root, or one whose intermediate signature has been tampered with. We
validated each with a proof-of-concept chain run against OpenSSL as a reference
oracle, on the affected stable releases through v5.9.1. Under our threat model this
is an unconditional vulnerability: the attacker supplies only the certificate chain
a verifier already accepts from a peer and holds no keys, and the bypass occurs with
the respective API used exactly as documented and no non-default configuration,
so a wolfSSL client could be made to authenticate a chain that reaches no trust
anchor. We
reported the issue to wolfSSL, who
acknowledged and reproduced it within roughly ninety minutes and escalated it to
top priority; the maintainers developed a patch based on our proposed fix, which we
confirmed the next day closed the bypass; the fix shipped in wolfSSL~5.9.2,
with two CVEs assigned:
CVE-2026-11310 and CVE-2026-11999. Given
the severity, we requested and received the maintainers' permission to share the
vulnerability in this paper ahead of public disclosure.

\noindent\textbf{Security Bug: JWT Claim Validation.} \sys also found an
audience-validation bypass in a widely used PHP JWT library. A JWT's audience (\texttt{aud}) claim names the
recipients a token is meant for, and an application should accept a token only if
its own identifier appears in that claim; RFC~7519 defines \texttt{aud} as a string
or an array of strings. \sys found that this library accepts a malformed audience claim in a way that lets a token intended for one audience satisfy a check for another.

The cause is a type-confusion path between claim parsing and audience membership checking, where a malformed claim is coerced into a representation that the audience check treats as valid.

We confirmed that a token intended for one audience could be reshaped so that a second audience check succeeded. We classify
this as a library-level
security bug rather than an unconditional vulnerability, since exploitation requires
a trusted signing path that can emit attacker-shaped claims; it nevertheless shows
how an application that treats the audience as an authorization boundary could
accept a lower-privilege token at an admin-facing endpoint.

\noindent\textbf{Ambiguity: JWT Base64 Decoding.} \sys traced a Base64URL-decoding
ambiguity in JWT verification to a flaw in an open-source RAG system that allows
an attacker to bypass token revocation. A JWT is carried as a compact
string of Base64URL-encoded
segments, and RFC~7515 prescribes a canonical encoding with no extra characters; for
some segment lengths, however, the unused low-order bits in a segment's final
Base64URL character can be changed to yield a different string that decodes to the
same bytes. The system's JWT library verifies the decoded signature bytes and does
not require a canonical spelling, so two textually distinct tokens can verify as the
same signed token. Stricter decoders reject the non-canonical form, which leaves each
implementation defensible on its own; the hazard surfaces only in an application that
assumes the token string uniquely identifies the token.

In that application, revocation is keyed on the serialized token string, while verification accepts byte-equivalent encodings. The two notions of token identity diverge: a byte-equivalent spelling of a revoked token misses the string-keyed blacklist yet still verifies. The permissive decoding behavior is intentional in the underlying library for compatibility with existing tokens.

We confirmed the bypass against a running deployment: after a token was revoked, the system rejected its canonical spelling but still accepted a byte-equivalent spelling. A token with corrupted signature bytes was rejected throughout, so
the cause is non-unique serialization. This is an
ambiguity realized as a downstream vulnerability where an attacker
must already hold a valid token, but once its canonical spelling is revoked an
equivalent spelling stays valid until expiry, so logout and refresh-token rotation fail
to revoke every serialization of the token.

\noindent\textbf{Ambiguity: SAML Assertion Replay.} \sys traced a SAML
response-replay ambiguity to a widely deployed identity service. A signed SAML response proves that an assertion is authentic
but not that it is fresh, and stopping reuse requires stateful, one-time-use tracking of
accepted assertions that the specification pins to no single layer. SAML libraries
disagree on who provides it: some maintain a replay cache themselves, many only expose the assertion identifier and leave the storage and the check to the caller, and even those that can enforce it often leave it off until configured. The
guarantee is thus contingent, and at each layer it is tempting to assume the layer below
already provides it.

The identity service sits in the middle of exactly such a stack. Its SAML library delegates replay protection, while the applications above never see the assertion. One authentication path uses one-time-use state, while another accepts a signed response without equivalent replay tracking.

We confirmed the bypass against a running deployment: a single captured signed response could be redeemed repeatedly,
while a response with a tampered signature was rejected throughout. Here the ambiguity is one of
responsibility, surfacing as a downstream vulnerability where the attacker
must capture a victim's valid signed
identity-provider-initiated response, after which it can be redeemed again.
 \section{Discussion}

We demonstrate \sys on cryptographic protocols since they meet its enabling conditions by construction (Section~\ref{sec:amplification}), but the idea need not stop there. Organizing discovery around a shared interface lets one unit of compute search many implementations at once, turning effort spent on a single target into coverage across an ecosystem. Project-local auditing, which works through a codebase file-by-file or commit-by-commit, captures none of this leverage. We see value in future work that brings this style of amplification to other domains, even where the shared interface must be engineered rather than assumed. The tracing stage should generalize more readily, since it needs only a behavioral fact about a library and a graph of its dependents, and can amortize a finding from any discovery method. Lowering the per-issue cost of detection this way stands to make thorough auditing accessible to defenders, and to shift the economics against attackers who otherwise benefit when each target is analyzed afresh.

One challenge to scaling is that adherence to a protocol varies from one implementation to the next. The techniques work best when every implementation intends to follow the specification as closely as possible, so that a disagreement points to a meaningful difference. Implementations that do not enforce strictness behave less predictably and add noise to the differential. This makes them easy to overlook when choosing what to harness, even though a permissive implementation may warrant the most scrutiny.

A recurring pattern in our findings is that libraries delegate security-critical checks, such as SAML replay protection or JWT revocation, to the application above them, often without documenting that they do. Applications then misread successful verification as trusted verification. We encourage maintainers to design for misuse resistance, adopting safer defaults that close these hazards rather than deferring them upward. A complementary direction is a formal catalog of the security properties each library leaves unenforced, which a dependency-aware tool, in the spirit of Dependabot,\footnote{\url{https://github.com/dependabot}} could surface automatically so that applications know which guarantees they remain responsible for.

 \section{Related Work}
\label{sec:related}
  \noindent\textbf{AI-assisted discovery:} 
  A growing body of work utilizes language models for fuzzing and vulnerability analysis. 
  LLMs have proven useful due to their ability to ingest or internalize large amounts of knowledge, as demonstrated by mGPTFuzz~\cite{maOneThousandPages2024}.
  Often, these are combined with dynamic execution and LLM-supplied fuzzing or test cases, such as in ChatAFL~\cite{mengLargeLanguageModel2024a}, which constructs a grammar to aid LLM fuzzing, LLMIF~\cite{wangLLMIFAugmentedLarge2024a}, which utilizes spec-aided LLM fuzzing, ProtocolGuard~\cite{ProtocolGuardDetectingProtocol}, which creates LLM assertions and LLM-supplied test cases, and PromptFuzz~\cite{lyuPromptFuzzingFuzz2024}, which uses LLMs to instrument fuzzers. \sys differentiates itself from these by analyzing libraries for specification ambiguity and cascading those to downstream apps rather than simply ensuring adherence to a potentially ambiguous specification. 
  Another similar work, GPTAid~\cite{liuGeneratingAPIParameter2025}, creates security rules for API calls and refines them over time. \sys instead relies on a deterministic verification engine.  Further, GPTAid largely focuses on runtime errors, while \sys surfaces many logic-level issues with accept/reject differentials.

\noindent\textbf{Differential testing: } Differential testing is a long-standing tool for exposing bugs in software by treating cross-implementation disagreement as an oracle. Frankencerts~\cite{brubakerUsingFrankencertsAutomated2014,petsiosNEZHAEfficientDomainIndependent2017}
  generated synthetic X.509 certificates and compared certificate-validation differentials. 
HVLearn~\cite{sivakornHVLearnAutomatedBlackBox2017} scanned hostname verification across TLS libraries. Symbolic analysis of PKCS\#1 v1.5 signature verification identified 15 semantic flaws among implementations~\cite{chauAnalyzingSemanticCorrectness2019}. 
Token Time Bomb~\cite{yangTokenTimeBomb2026} introduces JWTeemo that performs protocol-specific differential testing of JWT implementations through a custom grammar.
 As a whole, these systems demonstrate the value of implementation disagreement for finding security bugs, but they are largely specialized to a single format, validation routine, or cryptographic primitive, and several depend on domain-specific details.

\noindent\textbf{Boundary failures: } Many cryptographic and authentication vulnerabilities arise from the boundaries between libraries, frameworks, SDKs, and applications. SSL certificate validation, for example, has been shown to be severely misused in non-browser applications~\cite{georgievMostDangerousCode2012,somorovskyBreakingSAMLBe2012}. 
 Empirical analyses of OAuth SSO deployments further support this~\cite{sunDevilImplementationDetails2012}.
  Related work demonstrates that SDK assumptions, protocol repurposing, and relying-party logic elicit exploitable downstream failures~\cite{wangExplicatingSDKsUncovering2013,chenOAuthDemystifiedMobile2014,zhouSSOScanAutomatedTesting2014}. 
  Broader prevalence studies confirm that
  cryptographic API misuse is pervasive,
  with up to 88\% of Android apps using cryptographic APIs making at least one
  security-relevant
  mistake~\cite{fahlWhyEveMallory2012a,egeleEmpiricalStudyCryptographic2013a}. These
  findings establish that
  downstream applications routinely assume security from lower levels without checking deeply. \sys builds on this insight: rather than cataloging misuse
  patterns after the fact, it first identifies library- or spec-level ambiguities, then audits downstream applications for failing to account for these ambiguities.

  \noindent\textbf{Downstream impact.} Package ecosystems
  concentrate risk~\cite{zimmermannSmallWorldHigh2019}.
  Recent work has moved
  toward more precise
  impact analysis: function-level inter-package call graphs have been used to
  reduce false positives in npm vulnerability
  reports~\cite{puNoiseSignalPrecisely2026}, clone detection and reachability
  analysis have been applied to C/C++ supply
  chains~\cite{Hu_2026_VulSCA}, and
  ChainFuzz~\cite{dengChainFuzzExploitingUpstream2025a} generates downstream
  proof-of-concept exploits for known upstream vulnerabilities.
Brahmastra~\cite{bhoraskarBrahmastraDrivingApps2014} demonstrated
  component-first downstream testing by driving Android apps to exercise vulnerable
  third-party components. ZipDiff~\cite{youMyZIPIsnt2025} demonstrates vulnerability cases where downstream applications parse inputs with divergent libraries. A core feature of all these works is that they depend on a documented upstream vulnerability. \sys generalizes this pattern to any upstream ambiguity.
 \section{Conclusion}
We presented \sys, a system for discovering cryptographic misuse vulnerabilities
that pairs differential testing with AI so each covers the other's weakness:
differential testing supplies a verifiable signal that a discrepancy exists, while
agents contribute the protocol knowledge and adaptive search needed to drive
implementations into meaningful disagreement without hand-built grammars. Building
on this signal, discrepancy tracing inverts the prevailing paradigm of agentic
discovery: rather than audit one codebase for many flaws, \sys catalogs a flaw once
at the library level and propagates it across a cryptographic dependency graph to
the applications that inherit it.
Across X.509, JWT, and SAML, spanning 47 libraries and 8 languages, \sys surfaced
over 100 findings, including a critical chain-validation bypass in
wolfSSL, whose TLS stack ships in billions of devices and which its maintainers
confirmed and patched within hours of submission. The approach generalizes beyond
cryptographic protocols wherever a shared interface lets one unit of compute search
many implementations at once, shifting the economics of auditing toward defenders.

\appendices
\section{Supplementary Materials}

\newpage
\subsection{Blind Agent Prompt}

\begin{figure}[h]
  \begin{minipage}{\linewidth}
    \begin{lstlisting}[
      style=CStyleSmall,
      breaklines,
      frame=single,
      caption={Blind wolfSSL agent prompt.},
      captionpos=b,
      label=listing:wolfssl-blind-prompt
    ]
You are running on a dedicated sandbox VM with full filesystem access.

Task: investigate whether the wolfSSL checkout has a certificate authentication or X.509 validation bypass related to `src/x509_str.c`.

Rules:
- Use only the assigned wolfSSL checkout and your assigned artifact directory.
- Do not use internet sources.
- Treat the repository source and locally generated tests as the source of truth.
- Do not modify the wolfSSL checkout except through temporary build products inside your assigned artifact directory.

Required output:
- Create all artifacts under the assigned artifact directory.
- If you find a candidate issue, build and run a concrete validation.
- If the candidate does not reproduce, say that clearly.
- Leave rerunnable artifacts, including source/harness or scripts, a build/run script, `run.log`, and `NOTES.md`.
- `NOTES.md` must include the commit tested, commands to rerun, observed result, conclusion, and any limitations.
- At the end, print a concise final report with the artifact path and verdict.

Your goal is to determine whether there is a real, validated bypass on this checkout without any hints beyond the file/theme above.

Assigned artifact directory: `/home/ubuntu/experiment-batches/batches/wolfssl-blind-main-20260609-034627/claude-03/artifacts`
    \end{lstlisting}
  \end{minipage}
\end{figure}
\clearpage

\refstepcounter{subsection}
\addcontentsline{toc}{subsection}{\protect\numberline{\thesubsection}Detailed Capability Heatmaps}
\label{app:detailed-capability-heatmaps}

\begin{figure*}[t]
  \centering
  {\large\bfseries \thesubsection\quad Detailed Capability Heatmaps\par}
  \vspace{0.75em}
  \includegraphics[width=.98\textwidth]{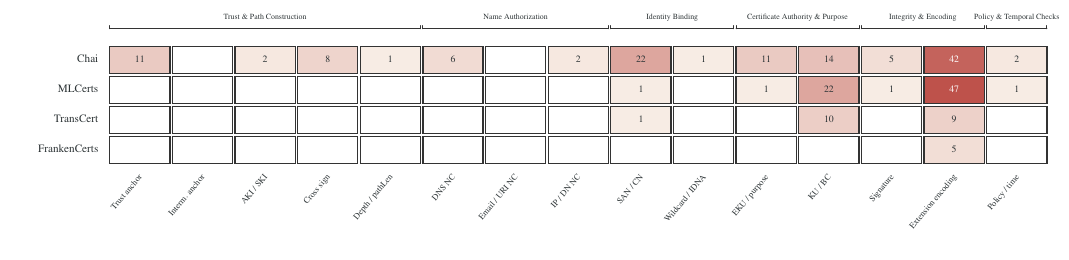}\\[.55em]
  \includegraphics[width=.98\textwidth]{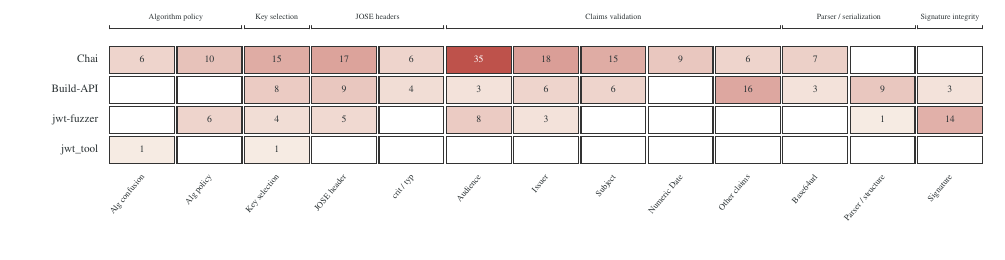}\\[.55em]
  \includegraphics[width=.98\textwidth]{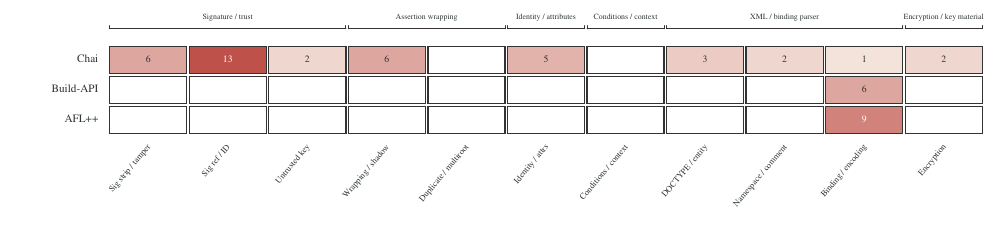}
  \caption{Detailed capability coverage heatmaps for X.509, JWT, and SAML.}
  \label{fig:capability-heatmaps-detailed}
\end{figure*}
 \clearpage

\bibliographystyle{IEEEtran}

\end{document}